\documentclass[reprint, superscriptaddress, prx]{revtex4-2}

\usepackage{siunitx}
\usepackage[english]{babel}
\usepackage[utf8]{inputenc}

\usepackage{float}
\usepackage{graphicx}   

\usepackage{amsmath}
\usepackage{amsfonts}
\usepackage{dsfont}
\usepackage{relsize}
\usepackage{braket}
\usepackage{bbm}
\usepackage{bm}
\usepackage{cancel}

\usepackage[section]{placeins}
\usepackage{xcolor}
\usepackage{layouts}
\usepackage{setspace}

\usepackage[colorlinks, hypertexnames=false]{hyperref}
\definecolor{navygray}{RGB}{110,140,170}
\hypersetup{
citecolor={navygray},
linkcolor={navygray},
urlcolor={navygray},
}
\usepackage[all]{hypcap}

\newcommand{\rcite}{Ref.~\cite}

\newcommand{\eref}[1]{\hyperref[#1]{{Eq.~\ref{#1}}}}  
\newcommand{\eqsref}[1]{\hyperref[#1]{{Eqs.~\ref{#1}}}}  

\newcommand{\fref}[1]{\hyperref[#1]{{Fig.~\ref{#1}}}}
\newcommand{\frefadd}[2]{\hyperref[#1]{{Fig.~\ref*{#1}#2}}}

\newcommand{\sref}[1]{\hyperref[#1]{{Sec.~\ref{#1}}}}
\newcommand{\aref}[1]{\hyperref[#1]{{App.~\ref{#1}}}}

\DeclareFontEncoding{LS1}{}{}
\DeclareFontSubstitution{LS1}{stix}{m}{n}
\DeclareSymbolFont{stixletters}{LS1}{stix}{m}{it}
\DeclareMathAccent{\cev}{\mathord}{stixletters}{"91}
\DeclareMathAccent{\vec}{\mathord}{stixletters}{"92}
\DeclareMathAccent{\vecev}{\mathord}{stixletters}{"95}

\begin{document}
\title{Solomon~equations~for~qubit~and~two-level~systems: \\ Insights~into~non-Poissonian~quantum~jumps}

\author{Martin Spiecker}
\email{martin.spiecker@kit.edu}
\affiliation{IQMT,~Karlsruhe~Institute~of~Technology,~76131~Karlsruhe,~Germany}
\affiliation{PHI,~Karlsruhe~Institute~of~Technology,~76131~Karlsruhe,~Germany}

\author{Andrei I. Pavlov}
\affiliation{IQMT,~Karlsruhe~Institute~of~Technology,~76131~Karlsruhe,~Germany}

\author{Alexander Shnirman}
\affiliation{IQMT,~Karlsruhe~Institute~of~Technology,~76131~Karlsruhe,~Germany}
\affiliation{TKM,~Karlsruhe~Institute~of~Technology,~76131~Karlsruhe,~Germany}

\author{Ioan M. Pop}
\email{ioan.pop@kit.edu}
\affiliation{IQMT,~Karlsruhe~Institute~of~Technology,~76131~Karlsruhe,~Germany}
\affiliation{PHI,~Karlsruhe~Institute~of~Technology,~76131~Karlsruhe,~Germany}
\affiliation{PI1,~Stuttgart~University,~70569~Stuttgart,~Germany}

\date{\today}

\begin{abstract}
We measure and model the combined relaxation of a qubit coupled to a discrete two-level system~(TLS) environment, also known as the central spin model. If the TLSs are much longer-lived than the qubit, non-exponential relaxation and non-Poissonian quantum jumps can be observed. In the limit of large numbers of TLSs, the relaxation is likely to follow a power law, which we confirm with measurements on a superconducting fluxonium qubit. Moreover, the observed relaxation and quantum jump statistics are described by the Solomon equations, for which we present a derivation starting from the general Lindblad equation for an arbitrary number of TLSs. We also show how to reproduce the non-Poissonian quantum jump statistics using a diffusive stochastic Schrödinger equation.
The fact that the measured quantum jump statistics can be reproduced by the Solomon equations, which ignore the quantum measurement backaction, hints at a quantum-to-classical transition.
\end{abstract}

\maketitle

\section{Introduction} \label{sec_intro}
Relaxation processes induced by spin environments are not only encountered in many areas of experimental physics, but also in theoretical physics as a popular toy model for understanding decoherence processes and the crossover from the quantum to the classical world \cite{Bloembergen1948Apr, deLange2010, Zhu2011Oct, Kubo2011Nov, Chekhovich2013Jun, Broadway2018Mar, Abobeih2018Jun, Abobeih2019Dec, Dasari2022Dec, Taylor2003Dec, Barnes2012, Jing2018, Zurek1982, Blume-Kohout2005Nov}.
A thorough understanding of spin relaxation was first achieved by studying long-lived nuclear spins in the field of nuclear magnetic resonance (NMR). 
One of the most important NMR spectroscopy methods for the structural analysis of molecules and even proteins \cite{Borgias1990Jan, Vogeli2014Apr, friebolin2010}
is based on the nuclear Overhauser effect and its description via the Solomon equations \cite{Solomon1955Jul}.
Here, we use the Solomon equations on a physically very different but conceptually similar spin system, consisting of a superconducting qubit coupled to a collection of two-level systems (TLSs).

In the field of superconducting quantum hardware, a common source of decoherence can be attributed to different types of TLSs that are typically short-lived compared to the qubit and therefore provide a well-thermalized environment \cite{Lisenfeld2016, Cywinski2009}. However, in a recent experiment, the interaction with long-lived TLSs, potentially electronic spins, has been reported \cite{Spiecker}. In view of these experimental findings and similar observations of memory effects in the qubit environment \cite{Gustavsson2016Dec}, we present here a derivation of the Solomon equations for an arbitrary number of TLSs, starting from a general Lindblad equation \cite{Lindblad1976Jun, Gorini1976May} for the qubit and the TLSs. 

\begin{figure}[t!]
\begin{center}
\includegraphics[scale=1.0]{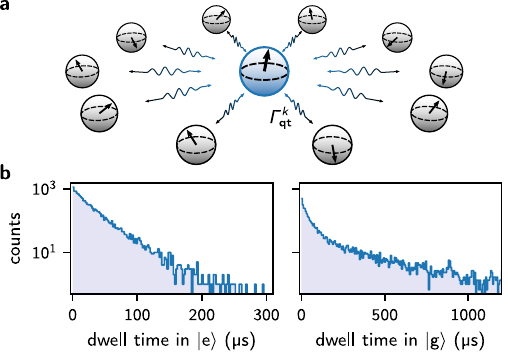}
\caption{\textbf{Qubit and long-lived TLSs.} \textbf{a} The qubit acts as the central spin for the surrounding TLSs. Under the presence of decoherence, their interaction leads to a mutual cross-relaxation with individual rates $\Gamma_\text{qt}^{k}$. \textbf{b} Measured non-exponential quantum jump statistics of our fluxonium qubit dwell times in the excited state ($|\text{e}\rangle$, left panel) and the ground state ($|\text{g}\rangle$, right panel). The statistics is based on 12000 counts corresponding to an $\sim\SI{4}{s}$ quantum jump trace. The bin size is \SI{2}{\micro s}. For $|\text{g}\rangle$ the histogram was binned by a factor of 3 in order to reduce the noise. The device is the same as in \rcite{Spiecker}.} \label{fig_intro}
\end{center}
\end{figure}

Interestingly, in contrast to NMR, the superconducting qubits and the TLSs are of a very different physical nature. Therefore, we can expect the qubit to interact with a large number of TLSs with potentially different frequencies, coupling strengths, and coherence times (\frefadd{fig_intro}{a}).
Consequently, we show how to deal with Solomon equations of infinite size for a given cross-relaxation distribution. We deduce the resulting power-law relaxation on long timescales, which we show to be in agreement with our experiments. 
Beyond superconducting devices, our analysis may also prove useful for the accurate description of dipolar relaxation processes, similar to \rcite{Staudenmaier2022Oct}. 
In addition, our results invite more NMR methods to be used on quantum processors, for instance using the nuclear Overhauser enhancement spectroscopy method to identify spurious qubit couplings.

Last but not least, we show that the measured non-Poissonian quantum jump statistics of our superconducting qubit (\frefadd{fig_intro}{b}) can be reproduced by a diffusive stochastic Schrödinger equation.
Surprisingly, the quantum jump statistics can also be generated in a much simpler way using the Solomon equations by essentially neglecting the measurement backaction on the TLSs. 
This procedure can therefore serve as a testbed to study the quantum-to-classical transition~\cite{Bassi2013Apr, Bethke2020Jul}. Our quantum jump analysis can readily be utilized on quantum processors with the TLSs replaced by qubits. 
We speculate that a reduced measurement backaction indicates the transition to a chaotic regime, as discussed in~\rcite{Berke2022May}, and could falsify error syndromes in quantum error-correction protocols.

We consider a mesoscopic superconducting qubit interacting with $n$ TLSs via a $\sigma^x$-type coupling, while the TLSs are assumed non-interacting. The system Hamiltonian is given by
\begin{align}
H = \frac{\hbar \omega_\text{q}}{2} \sigma^\text{q}_z + \sum_{k = 1}^n \frac{\hbar \omega_k}{2}\sigma^k_z +  \sum_{k = 1}^n \hbar g_k \sigma^\text{q}_x \sigma^k_x, \label{eq_hamiltonian}
\end{align}
where $\omega_\text{q}$ and $\omega_k$ are the frequencies of the qubit and the $k^\text{th}$ TLS, $g_k$ is their coupling strength and $\sigma^\text{q}_{x, z}$ and $\sigma^k_{x, z}$ denote their Pauli matrices.
The physics of the central spin model \cite{Prokofev2000, Hsieh2018, Villazon2020Aug} as of \eref{eq_hamiltonian} with added dissipation is generally complex and an active field of research, including phenomena such as superradiance \cite{Kessler2010, Shao2023}, state revivals \cite{Dooley2013May, He2019May, Fan2023}, dynamical phase transitions \cite{Alvarez2006, Pastawski2007, Marino2022Oct}, and exceptional points \cite{Tserkovnyak2020Jan}.
With increasing decoherence in the system, one can expect a transition from coherent oscillations to a regime where the expectation values of the populations follow a simple rate equation. This means that the qubit population $p_\text{q}$ should be governed by a linear differential equation of the form 
\begin{align}
\dot{p}_\text{q}(t) = - \Gamma_1(t) [p_\text{q}(t) - p_\text{eq}(t)]  \label{eq_relax_qubit}
\end{align}
with potentially time-dependent coefficients $\Gamma_1(t)$, the qubit relaxation rate, and $p_\text{eq}(t)$, the equilibrium population, i.e., the qubit population at which $\dot{p}_\text{q}(t)$ would vanish. The coefficients $\Gamma_1(t)$ and $p_\text{eq}(t)$ are given by the current state of the TLS environment, which in turn might also depend on the qubit history. 
The main task is therefore to validate \eref{eq_relax_qubit} and to derive exact or approximate solutions for different scenarios, which in retrospect allow us to draw conclusions on the TLS environment.

We want to note that by observing the qubit population dynamics alone, $\Gamma_1(t)$ and $p_\text{eq}(t)$ are \textit{a priori} unknown functions that cannot be disentangled. However, if one has access to both qubit transition rates $\Gamma_{\uparrow,\downarrow}$, for instance
by resetting the qubit to its ground or excited state and measuring the subsequent population change, then the time-dependent transition rates are
\begin{gather}
	\Gamma_\uparrow(t) = \dot{p}_\text{q}(t)\big|_{p_\text{q} = 0} \quad \text{and} \quad \Gamma_\downarrow(t) = - \dot{p}_\text{q}(t)\big|_{p_\text{q} = 1} \label{eq_def_transition_rates}
\end{gather}
after resetting to the ground or excited state, respectively. This method is applicable for any qubit with active state reset capability.
With \eref{eq_relax_qubit}, one then obtains the usual expressions
\begin{align}
	\Gamma_1(t) &= \Gamma_\uparrow(t) + \Gamma_\downarrow(t), \label{eq_def_relax_rate} \\
	p_\text{eq}(t) &= \Gamma_\uparrow(t) / \Gamma_1(t). \label{eq_def_peq}
\end{align}
For a TLS environment, it turns out that $\Gamma_1$ is constant over time, i.e., it does not depend on the TLS populations, which also implies that $\Gamma_1$ is temperature-independent (cf. \eqsref{eq_redfield_relaxation} and \ref{eq_relax_rates} and \rcite{Spiecker}).

The manuscript is structured as follows. We begin the discussion with a review of the Bloch-Redfield master equation, which is applicable for a large but weakly coupled TLS bath (\sref{sec_bloch_redfield}). In this case, the bath populations remain unaffected by the qubit dynamics, and consequently $p_\text{eq}$ is constant in time. We then transition to a countable set of TLSs, for which we derive the Solomon equations (\sref{sec_solomon}). Next, we apply the Solomon equations to the situation in which the TLSs are not interacting with each other and where the relaxation matrix has the shape of a so-called arrowhead matrix (\sref{sec_solomon_analytics}). We then present an analytical solution of the rate equation for the scenario in which all TLSs have the same cross-relaxation rate with the qubit. For the case of a cross-relaxation distribution, we show that the relaxation becomes a power-law on long timescales. We then discuss the qubit relaxation as a function of the mutual decoherence and show that the relaxation of the Bloch-Redfield master equation can be recovered (\sref{subs:T1_one_TLS}).
In a final step, we use the Solomon equations to simulate the qubit's non-Poissonian quantum jump statistics and show its agreement with our measurements (\sref{sec_quantum_jumps}). 

\section{Bloch-Redfield master equation} \label{sec_bloch_redfield}
For a later comparison to the Solomon equations, we briefly review the derivation and results of the widely used Bloch-Redfield master equation with secular approximation \cite{Bloch1957Feb, Redfield1957Jan}, also known as the Born-Markov or quantum optical master equation \cite{breuer2002theory}.
We closely follow the derivation and notations of Ref.~\cite{Schaller}. The Bloch-Redfield master equation rests on the Born approximation. With the qubit being the system and the TLSs forming the bath, the Born approximation is valid when the qubit interacts weakly with a large number of TLSs, in which case their populations will not be altered significantly by the qubit and can be approximated as constant.
In close relation, one assumes that there are no initial system bath correlations and that the bath itself is in an equilibrium state that does not evolve in time.
For a small number of TLSs, with a long intrinsic lifetime, the Born approximation is certainly not fulfilled, requiring a description with the Solomon equations (s. \sref{sec_solomon}).

The Bloch-Redfield master equation is typically derived in the interaction picture.  
Here, the qubit's reduced density matrix $\rho_\text{q}$ in second-order perturbation is governed by
\begin{multline}
\dot{\rho}_\text{q}(t) = - \frac{1}{\hbar^2} \int\limits_0^\infty\! \,C(\tau) [\sigma^\text{q}_x(t),\,\sigma^\text{q}_x(t - \tau) \rho_\text{q}(t)] \\[-1.0ex]
+ C^*(\tau) [\rho_\text{q}(t)\sigma^\text{q}_x(t - \tau),\,\sigma^\text{q}_x(t)]\,\text{d}\tau, \label{eq_redfield}
\end{multline}
where we have already used the first and second Markov approximations, i.e., evaluating the qubit's density matrix at the current time $t$ and extending the integration over the bath correlation function $C(\tau)$ to infinity.
With the secular approximation, which means removing the explicit time-dependence, the above equation becomes the Bloch-Redfield master equation, which 
is of Lindblad type.
\begin{figure}[t!]
\begin{center}
\includegraphics[scale=1.0]{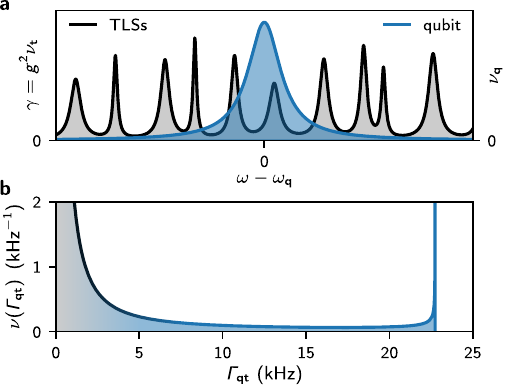}
\caption{\textbf{Illustrative model for a cross-relaxation distribution.} \textbf{a} Spectral coupling density $\gamma$ of the TLSs and density of states $\nu_\text{q}$ of the qubit. The TLSs are randomly distributed in frequency with individual coupling strength to the qubit and intrinsic decoherence.
\textbf{b} Density of cross-relaxation rates following from \eref{eq_gamma_qt} for a random TLS frequency distribution, assuming the same coupling strength $g$ and mutual decoherence $\Gamma_2$ for all TLSs. The parameters are taken from Ref. \cite{Spiecker}. If the TLSs have some distribution in $g$ and $\Gamma_2$, the divergence at the maximal cross-relaxation rate washes out and disappears, whereas the divergence for low cross-relaxation rates remains. The color gradient from blue to gray illustrates the increasing frequency detuning between the qubit and the TLSs of the corresponding cross-relaxation rates.} \label{fig_Gqt}
\end{center}
\end{figure}
The bath correlation function is given by
\begin{align}
C(\tau) &=  \hbar^2 \sum_k g_k^2 \,\text{Tr}\{\sigma^k_x(t) \sigma^k_x(t - \tau)\rho_\text{TLSs}\}  \nonumber \\
        &=  \hbar^2 \sum_k g_k^2 \left[ p_\text{t}^k e^{i \omega_k \tau} + (1 - p_\text{t}^k) e^{- i \omega_k \tau}\right], \label{eq_correlation_sum}
\end{align}
where $\rho_\text{TLSs}$ is the time-independent density matrix of the TLS bath, which contains the populations $p_\text{t}^k$ of the TLSs. Typically, the TLSs are assumed to be in thermal equilibrium and given by the Fermi-Dirac distribution $p_\text{t}^{k} = n(\omega_k)$. As a consequence, the qubit will relax to the same thermal equilibrium with $p_\text{q} = n(\omega_q)$. 

The usual way to continue the derivation from \eref{eq_correlation_sum} is to consider the limit where the TLSs become dense in frequency with a vanishing coupling strength such that they form a spectral coupling density $\gamma(\omega) = \langle g^2\nu_\text{t}\rangle(\omega)$, with $\nu_\text{t}$ being the TLS density \cite{Carmichael1999, Schaller}. Alternatively,
this transition to a density may be viewed as an ensemble average over qubit and TLS frequencies for many instances of the experiment. 
In \frefadd{fig_Gqt}{a} we illustrate $\gamma(\omega)$ and the qubit's density of states $\nu_\text{q}$.
The correlation function can now be expressed via the integral
\begin{align}
C(\tau) &= \hbar^2 \!\int_0^\infty \!\! \gamma(\omega) \left[ p_\text{t}(\omega) e^{i \omega \tau} + (1 - p_\text{t}(\omega)) e^{- i \omega \tau}\right] \text{d}\omega \nonumber \\
        &= \hbar^2 \!\int_{-\infty}^\infty [s(\omega) + a(\omega)] e^{i\omega \tau} \text{d}\omega, \label{eq_correlation_integral}
\end{align}
where we introduce the symmetric $s(\omega) = s(-\omega)$ and antisymmetric $a(\omega) = - a(- \omega)$ frequency components of the bath correlation function with
$s(\omega \geq 0) =\gamma(\omega) / 2$ and $a(\omega > 0) =\gamma(\omega) \cdot [p_\text{t}(\omega) - 1 / 2]$, which helps to compute the half-sided inverse Fourier transform in \eref{eq_redfield}. 

Next, we can rewrite \eref{eq_redfield} as $\dot{\rho}_\text{q} = \mathcal{L} \rho_\text{q}$ by introducing the Liouvillian superoperator $\mathcal{L}$. Separating $\mathcal{L}$ in its real and imaginary parts, we define $\dot{\rho}_\text{q} = \dot{\rho}_\text{q}^\text{relax} + \dot{\rho}_\text{q}^\text{Lamb} = \text{Re}(\mathcal{L})\cdot \rho_\text{q} + i \text{Im}(\mathcal{L}) \cdot \rho_\text{q}$.
The real-part describes the qubit relaxation and decoherence
\begin{align}
\frac{\dot{\rho}_\text{q}^\text{relax}}{2 \pi \gamma(\omega_\text{q})} = - \frac{1}{2}\begin{pmatrix}
2\rho_{00} & \rho_{01}\\ 
\rho_{10} &  - 2\rho_{00}
\end{pmatrix} 
+ p_\text{t}(\omega_\text{q}) \sigma_z.  \label{eq_redfield_relaxation}
\end{align}
Here, we obtain the important result that the qubit relaxation $\Gamma_1 = 2 \pi \gamma(\omega_\text{q})$ and decoherence $\Gamma_2 = \Gamma_1 / 2$ are independent of the TLS populations \cite{Weiss2011Nov}. For completeness, the imaginary part incorporates the Lamb shift
\begin{align}
\dot{\rho}_\text{q}^\text{Lamb} = i\,\,\text{P.V.}\!\!\!\int\limits_{-\infty}^\infty\frac{2 \omega_\text{q}s(\omega)}{\omega_\text{q}^2 - \omega^2}\text{d}\omega \cdot \begin{pmatrix}
0 & -\rho_{01}\\ 
\rho_{10} &  0
\end{pmatrix}, \label{eq_lamb_shift}
\end{align}
where the integral has to be evaluated by means of the Cauchy principal value. Again, there is no dependence on the TLS populations. 
Note that by integrating also over negative frequencies, contributions from counter-rotating waves are included in the Lamb shift. Thus the Lamb shift vanishes for a constant spectral coupling density.
If we had applied the rotating-wave approximation in the Hamiltonian in \eref{eq_hamiltonian}, the Lamb shift would not contain contributions from counter-rotating waves \cite{Agarwal1973Mar, Carmichael1999} and for a constant spectral coupling density the Lamb shift would diverge.

In view of the following section, the Bloch-Redfield master equation can also be used to estimate the cross-relaxation rate between the qubit and a single TLS. Assuming the qubit and TLS spectral densities are Lorentz distributions centered at $\omega_\text{q}$ and $\omega_\text{q} + \delta$ with linewidths $\Gamma^\text{q}_2$ and $\Gamma^\text{t}_2$, respectively (s. \frefadd{fig_Gqt}{a}),
then, for a short time such that the Born approximation remains valid, we can calculate the average qubit relaxation rate and with that obtain the cross-relaxation rate
\begin{align}
\Gamma_\text{qt} = \frac{2g^2\Gamma_2}{\Gamma_2^2 + \delta^2}, \label{eq_gamma_qt}
\end{align} 
with the mutual decoherence $\Gamma_2 = \Gamma^\text{q}_2 + \Gamma^\text{t}_2$. 
Expanding the scope to include multiple TLSs, we can now define the components of the total loss rate of the qubit $\Gamma_1 = \Gamma_\text{q} + \sum_k \Gamma_\text{qt}^k$. The rate $\Gamma_\text{q}$ accounts for additional losses from other environments.
If the TLS detunings $\delta_k$ are randomly distributed in frequency, \eref{eq_gamma_qt} defines the density of the cross-relaxation rates $\nu(\Gamma_\text{qt})$ depicted in \frefadd{fig_Gqt}{b}. 
Note that adding the rates implicitly assumes that we can ignore higher-order coherent effects in the system, as discussed in detail in the next section.

\section{Derivation of Solomon's equations} \label{sec_solomon}

For a small number of TLSs the Born approximation is no longer valid since their populations will be altered by the qubit excitation \cite{Esposito2003Dec, Pekola2023Jun}. Ideally, the qubit and the TLSs are treated on equal footing. 
In experiments, the dissipation required to yield a master equation is naturally provided by the surrounding environment of the qubit and the TLSs. In NMR, this environment is known as the spin-lattice for which a detailed derivation can be found in \rcite{Goldman2001Apr}. The spin-lattice can either be modeled as a classical noise source in the framework of the stochastic Liouville equation or treated quantum mechanically, essentially following the Bloch-Redfield formalism. Based on these results, we will construct the Liouvillian from a general Lindblad equation.

The general idea behind the Solomon equations is to reduce the Liouvillian to an equation of motion for the expectation values of the qubit and TLS populations~\cite{Solomon1955Jul, Kumar2000Sep}.
As will be discussed in more detail in the following, in case the decoherence and frequency spread in the system is large in comparison to the couplings, coherent effects can be neglected. An intermediate step is therefore the derivation of Pauli's master equation for the diagonal components of the density matrix. Here, we use the adiabatic elimination of the coherences. Alternatively, one may derive the Pauli master equation from a generalized Schrieffer-Wolff transformation~\cite{Kessler2012Jul}, yet others refer to it as the Nakajima-Zwanzig-Markov-Pauli master equation~\cite{Timm2011Mar}.
However, beyond the qubit and a single TLS, deriving closed-form solutions is practically infeasible due to the emergence of multispin phenomena (cross correlations), which are discussed in NMR within the framework of magnetization modes \cite{Kumar2000Sep}.

In this context, our main focus is to present a compact derivation of the Solomon equations, which we show to follow from the Markov approximations for the coherences and from the neglect of higher-order coherent processes. We accompany our derivation by showing explicit calculations for the case of a qubit coupled to a single TLS.
We formulate the Lindblad equation for the Hamiltonian in \eref{eq_hamiltonian} 
by including individual dissipators for the qubit and each of the TLSs:
\begin{align}
	\dot{\rho} = -\frac{i}{\hbar} [H, \rho] + \sum_\alpha L_\alpha \rho  L_\alpha^\dagger - \frac{1}{2} \{ L_\alpha^\dagger  L_\alpha, \rho\} .\label{eq_liouville_lindblad}
\end{align} 
Here, $L_\alpha$ are the jump operators, which are given by $\sqrt{\Gamma_\downarrow^j}\cdot\sigma_j^-$ and $\sqrt{\Gamma^j_\uparrow}\cdot\sigma_j^+$ for energy relaxation and excitation, respectively, and \raisebox{0pt}[8pt][7pt]{$\sqrt{\Gamma^j_\uparrow / 2}\cdot\sigma^j_z$} for dephasing. We use the index $j$ to denote both the qubit and the TLS variables.
Casting the Liouville-von Neumann equation (\eref{eq_liouville_lindblad}) in superoperator notation $\dot{\rho} = \mathcal{L} \rho$, the Liouvillian  $\mathcal{L}$ can be represented by a matrix and $\rho$ by a vector (s. \aref{sec_liouvillian}).
In view of Nakajima-Zwanzig's projection operator technique \cite{Nakajima1958Dec, Zwanzig1960Nov}, we may sort the density matrix $\rho = \begin{pmatrix}\rho_\text{D},\;\rho_\text{C} \end{pmatrix}^T$ for diagonal entries $\rho_\text{D}$ (populations) and non-diagonal entries $\rho_\text{C}$ (coherences). Then  \eref{eq_liouville_lindblad} reads
\begin{align}
	\dot{\rho} = \begin{pmatrix} 
		\mathbf{\Gamma} & \mathbf{R}^T \\
		\mathbf{R}& \mathbf{C} 
	\end{pmatrix} \rho.  \label{eq_Lindbladmatrix}
\end{align}
The matrix $\mathbf{\Gamma}$ depends only on the decay rates $\Gamma^j_{\uparrow,\downarrow}$ and describes the relaxation of the system to thermal equilibrium. 
The matrix $\mathbf{R}$ couples populations and coherences, giving rise to vacuum Rabi oscillations and energy exchange between the qubit and the TLSs. Therefore, its matrix elements $\mathbf{R}_{mn}$ are either zero or amount to one of the coupling terms, in short $\mathbf{R}_{mn} \in \{0, \pm i g_k\}$. The matrix $\mathbf{C}$ describes the oscillations, decoherence, and interference of the coherences between the qubit and TLSs as well as between the TLSs. Since the $\sigma^x$-coupling in \eref{eq_hamiltonian} induces only one-photon (flip-flop) and two-photon transitions (flip-flip), we can order the coherences $\rho_\text{C} = \begin{pmatrix}
    \rho_{\text{C}_\text{Z}},& \rho_{\text{C}_\text{D}},& \rho_{\text{C}_\text{R}},& \rho_{\text{Q}}
\end{pmatrix}^T\!$ and bring \eref{eq_Lindbladmatrix} into~the~form
\begin{align}
	\dot{\rho} = \begin{pmatrix} 
		\mathbf{\Gamma} & \mathbf{R}^T_\text{Z} & \mathbf{R}^T_\text{D} & 0 & 0\\[2pt]
		\mathbf{R}_\text{Z} & \mathbf{C}_\text{Z} & 0 & \mathbf{S}_\text{Z}^T & 0\\[2pt]
		\mathbf{R}_\text{D} & 0 &\mathbf{C}_\text{D} & \mathbf{S}_\text{D}^T & 0\\[2pt]
		0 & \mathbf{S}_\text{Z} & \mathbf{S}_\text{D} &  \mathbf{C}_\text{R} & 0\\[2pt]
		0 & 0 & 0 & 0 &\mathbf{Q} 
	\end{pmatrix} \, \rho  \label{eq_Lindbladmatrix_2}
\end{align}
with the indices Z and D denoting those entries that give rise to zero and double quantum transitions, respectively, i.e., excitation differences. The dynamics of the remaining coherences that do not directly couple to populations are described by $\mathbf{C}_\text{R}$ and couple via $\mathbf{S}_\text{Z,D}$ to the relevant coherences. Similar to $\mathbf{R}$, one finds $\mathbf{S}_{\text{Z,D}; mn} \in \{0, \pm ig_k\}$.
All irrelevant coherences $\rho_\text{Q}$ between even and odd photon states undergo an independent evolution described by $\mathbf{Q}$. The matrix $\mathbf{Q}$ makes up half the size of the Liouvillian.
For the qubit and a single TLS, $\mathbf{C}_\text{R}$ and $\mathbf{S}_\text{Z,D}$ do not exist and a transformation of the Liouvillian leads to the so-called Redfield kite~\cite{ernst1987principles, Kumar2000Sep}. The matrix structure of the Liouvillian is depicted in \aref{sec_liouvillian} as an illustration for the qubit coupled to one or two TLSs.

We may further decompose $\mathbf{C}_\text{Z,D} = \mathbf{D}_\text{Z,D} - \mathbf{O}_\text{Z,D}$ with $\mathbf{D}$ being diagonal and in charge of the oscillation and decoherence of the relevant coherences. The entries $\mathbf{D}_{\text{Z,D};mm}$, e.g. describing the coherence of the qubit with the $k^\text{th}$ TLS, are of the form $\mathbf{D}_{\text{Z};mm} = - \Gamma_2^k \pm i \delta_k$ and $\mathbf{D}_{\text{D};mm} = - \Gamma_2^k \pm i \sigma_k$, with the mutual decoherence $\Gamma_2^k = \Gamma_2^\text{q} + \Gamma_2^{\text{t}_k}$ and the general detunings $\delta_k = \omega_k - \omega_\text{q}$ and $\sigma_k = \omega_\text{q} + \omega_k$. The decoherence of the qubit $\Gamma_2^\text{q}$ and of the TLSs $\Gamma_2^{\text{t}_k}$ are as usual of the form $\Gamma_2 = \Gamma_\varphi + \Gamma_1 / 2$ with $\Gamma_1 = \Gamma_\uparrow + \Gamma_\downarrow$.
In case the qubit is coupled to only one TLS, the matrix $\mathbf{O}_\text{Z,D}$ is zero. For more TLSs, the coherences of different photon manifolds can interfere when the TLSs undergo relaxation processes \cite{Muller1987Nov, Kowalewski2017Dec}.
Thus, $\mathbf{O}_{\text{Z,D};mn} \in \{0, \pm\Gamma^{\text{t}_k}_\uparrow, \pm\Gamma^{\text{t}_k}_\downarrow\}$ (\aref{sec_liouvillian}), and in case the TLSs are lossless, the matrices $\mathbf{C}_\text{Z,D}$ are diagonal. 
For completeness, even though not applicable in \rcite{Spiecker}, in the scenario where the TLSs are coupled with each other, there are additional off-diagonal matrix elements that arise from vacuum Rabi oscillations. In this case, one has $\mathbf{O}_{\text{Z,D};mn} \in \{0, \pm\Gamma^{\text{t}_k}_\uparrow, \pm \Gamma^{\text{t}_k}_\downarrow, \pm ig_k\}$.
Lastly, it is important to mention that the diagonal entries of $\mathbf{C}_\text{R}$ always contain various combinations of decoherences and detunings.

In the case of a qubit and a single TLS, we find for the density matrix entries $\rho_{mn} = \langle v_m|\rho|v_n\rangle$ of the single-photon manifold \cite{Barends2013Aug}:
\begin{align} \label{eq_single_photon_manifold}
	\dot{\rho}_{11} &= ig (\rho_{12} - \rho_{21}) - (\Gamma^\text{t}_\downarrow + \Gamma^\text{q}_\uparrow) \rho_{11} + \Gamma^\text{q}_\downarrow \rho_{00} + \Gamma^\text{t}_\uparrow\rho_{33}, \nonumber  \\
	\dot{\rho}_{21} &= ig (\rho_{22} - \rho_{11}) - \left(\Gamma^\text{q}_2 + \Gamma^\text{t}_2 - i \delta \right) \rho_{21},\\
	\dot{\rho}_{22} &= ig (\rho_{21} - \rho_{12}) - (\Gamma^\text{q}_\downarrow + \Gamma^\text{t}_\uparrow) \rho_{22} + \Gamma^\text{t}_\downarrow \rho_{00} + \Gamma^\text{q}_\uparrow\rho_{33}, \nonumber
\end{align}
and $\rho_{12} = \rho_{21}^*$. The wave functions $\ket{v_0}$, $\ket{v_1}$, $\ket{v_2}$, $\ket{v_3}$ correspond to the states $\ket{11}$,  $\ket{01}$, $\ket{10}$, $\ket{00}$, respectively, with the first entry for the qubit and the second for the TLS. For more details on the notation, see \aref{sec_liouvillian}. In case the mutual decoherence $\Gamma_2 = \Gamma^\text{q}_2 + \Gamma^\text{t}_2$ is large compared to the timescales on which the occupations $\rho_{11}$ and $\rho_{22}$ vary, one may approximate $\rho_{21}$ by the population-dependent steady-state value, which leads to overdamped vacuum Rabi oscillations (\aref{sec_stochastic_liouville}). In view of \eref{eq_Lindbladmatrix} this means 
\begin{align}
	\rho_\text{C} = - \mathbf{C}^{-1} \mathbf{R} \cdot \rho_\text{D}. \label{eq_rhoCofD}
\end{align}
\indent This approximation is best understood in the framework of Nakajima-Zwanzig's master equation \cite{Nakajima1958Dec, Zwanzig1960Nov}. Solving the dynamics of the coherences in \eref{eq_Lindbladmatrix} gives
\begin{align}
	\rho_\text{C}(t) = e^{\mathbf{C}(t - t_0)} \rho_\text{C}(t_0) + \int_0^{t - t_0} \!\!\! e^{\mathbf{C} \tau} \mathbf{R}\,\rho_\text{D}(t - \tau) \text{d}\tau.
\end{align}
Insertion in the dynamics of the populations results in the integrodifferential equation
\begin{multline}
	\dot{\rho}_\text{D}(t) = \mathbf{\Gamma} \rho_\text{D}(t) + \mathbf{R}^T e^{\mathbf{C}(t - t_0)} \rho_\text{C}(t_0) \\ + \int_0^{t - t_0} \mathbf{R}^T e^{\mathbf{C} \tau}\,\textbf{}\mathbf{R}\,\rho_\text{D}(t - \tau) \text{d}\tau.
\end{multline}
Under the assumption that the populations vary slowly compared to the relaxation of the coherences, we can perform several approximations. First of all, we can neglect the second term as we are only interested in the slow population dynamics on long timescales. In the literature, it is often assumed instead, under the Born approximation, that the coupling between system and environment is switched on at $t_0$, in which case the initial coherences vanish on average \cite{Carmichael1999}. If the coupling is always on, the initial coherences  $\rho_\text{C}(t_0)$ do not necessarily vanish in thermal equilibrium (cf. \eref{eq_rhoCofD}). Next, we can also apply the first and second Markov approximations, as we did for the Bloch-Redfield equation (s. \sref{sec_bloch_redfield}). This means we can approximate $\rho_\text{D}$ by its current time $\rho_\text{D}(t - \tau) \approx \rho_\text{D}(t)$ and extend the integration to infinity, yielding
\begin{align}
	\dot{\rho}_\text{D}(t) &= \mathbf{\Gamma} \rho_\text{D}(t) + \mathbf{R}^T \mathbf{C}^{-1} e^{\mathbf{C} \tau}\mathbf{R} \Big|_0^\infty \rho_\text{D}(t)\nonumber \\
	&= \left(\mathbf{\Gamma} - \mathbf{R}^T \mathbf{C}^{-1} \mathbf{R}\right) \rho_\text{D}(t) = \mathcal{L}_\text{D} \, \rho_\text{D}(t). \label{eq_liou_relax}
\end{align}
This proves that \eref{eq_rhoCofD} follows from the two Markov approximations. For a quantitative discussion on the Markov approximations, see \aref{sec_stochastic_liouville}.

If the qubit is coupled to more than one TLS, calculating the inverse of $\mathbf{C}$ is not feasible. However, since we assumed for the Markov approximation that the coherences vanish rapidly, the diagonal entries of $\mathbf{C}$ must be large in comparison to the off-diagonal entries. Therefore, we can approximate the inverse up to second order in the diagonal entries (s. \aref{sec_matrix_inverse}). If the TLSs are lossless (and not interacting with each other) the matrices $\mathbf{C}_\text{Z,D}$ are diagonal and the inverse can readily be computed. Otherwise, we need to approximate $\mathbf{C}_\text{Z,D}$ to be diagonal (\aref{sec_matrix_inverse}) to yield the Solomon equations. We have
\begin{align}
    \mathbf{R}^T \mathbf{C}^{-1} \mathbf{R} &\approx \mathbf{R}_\text{Z}^T \mathbf{C}_\text{Z}^{-1} \mathbf{R}_\text{Z} + \mathbf{R}_\text{D}^T \mathbf{C}_\text{D}^{-1} \mathbf{R}_\text{D}   \label{eq_inverse_approx_1} \\
    &\simeq \mathbf{R}_\text{Z}^T \mathbf{D}_\text{Z}^{-1} \mathbf{R}_\text{Z} + \mathbf{R}_\text{D}^T \mathbf{D}_\text{D}^{-1} \mathbf{R}_\text{D}.
    \label{eq_inverse_approx_2}
\end{align}
Depending on the situation, one finds up to first or second order the cross-relaxation rates $\Gamma_\text{qt}^{\Delta_k}$ between the qubit and the $k^\text{th}$ TLS.
The cross-relaxation rates are of the form \cite{Solomon1955Jul, Barends2013Aug}
\begin{align} 
	\Gamma_\text{qt}^{\Delta_k} = \frac{2g_k^2 \Gamma_2^k}{(\Gamma_2^k)^2 + \Delta_k^2} \label{eq_cross_relaxation}
\end{align} 
with $\Delta_k \in \{ \delta_k, \sigma_k\}$. As a reminder, $\Gamma_2^k$ is the sum of the qubit and TLS decoherence rates. The derivation of the formula can be tracked in \eref{eq_single_photon_manifold} when the coherence $\rho_{12}$ is approximated by its population-dependent steady-state value.
In the case of a qubit and a single TLS, where none of the previously discussed approximations are needed, the relaxation matrix can be written as
\begin{align}
	\mathcal{L}_\text{D} &= \begin{pmatrix}
		- \Gamma^\text{q}_\downarrow - \Gamma^\text{t}_\downarrow & \Gamma^\text{q}_\uparrow & \Gamma^\text{t}_\uparrow & 0  \\
		\Gamma^\text{q}_\downarrow &  - \Gamma^\text{q}_\uparrow - \Gamma^\text{t}_\downarrow & 0 & \Gamma^\text{t}_\uparrow  \\
		\Gamma^\text{t}_\downarrow & 0 & -\Gamma^\text{q}_\downarrow - \Gamma^\text{t}_\uparrow & \Gamma^\text{q}_\uparrow \\
		0& \Gamma^\text{t}_\downarrow & \Gamma^\text{q}_\downarrow & - \Gamma^\text{q}_\uparrow - \Gamma^\text{t}_\uparrow
	\end{pmatrix} \nonumber \\
	&\qquad \qquad + \begin{pmatrix}
		-\Gamma^\sigma_\text{qt} & & & \Gamma^\sigma_\text{qt} \\
		&- \Gamma^\delta_\text{qt} & \Gamma^\delta_\text{qt} & \\
		&\Gamma^\delta_\text{qt} & - \Gamma^\delta_\text{qt} & \\
		\Gamma^\sigma_\text{qt} &  & & - \Gamma^\sigma_\text{qt} \\ 
	\end{pmatrix}
 \label{eq_lindbladian_transf}
\end{align}
and similarly for more TLSs (s. \aref{sec_proof_solomon}).

So far, we were able to reduce the $4^{n + 1}$-dimensional Liouvillian in \eref{eq_Lindbladmatrix} for the qubit and $n$ TLSs to a $2^{n + 1}$-dimensional rate equation for the occupations. However, we want to reduce the rate equation even further to the $(n + 1)$-dimensional Solomon equations, describing the expectation values of the populations, i.e., the measurable population probabilities $p_j$. In general, one cannot expect that there is an equation of motion for a closed set of expectation values, which is why several truncation strategies exist in the literature \cite{Leymann2014Feb, Roehling2018, Lindberg1994, Gartner2011, Fauseweh2017}. As we elaborate in the next paragraph, the approximations in \eqsref{eq_inverse_approx_1} and \ref{eq_inverse_approx_2} are sufficient to obtain the  Solomon equations for an arbitrary number of TLSs, which is the central result of this derivation.
For physical intuition, one can think of the approximation in \eref{eq_inverse_approx_1} as neglecting coherence between TLSs, i.e., the TLSs are not cooperative, and in \eref{eq_inverse_approx_2} as ignoring interference between different cross-relaxation pathways \cite{Muller1987Nov}.

The Solomon equations were originally derived for two nuclear spins, analogous to a qubit with a single TLS. In this case the derivation is straightforward and was first presented by Solomon \cite{Solomon1955Jul}. The general idea is to find a coordinate transformation $\mathbf{S}$ for $\rho_\text{D}$ that leads to a covariant description of the probabilities. The probabilities are obtained by computing the corresponding partial traces.
With our choice of ordering the diagonal entries (\aref{sec_liouvillian}), it holds that
\begin{gather*}
	p_\text{q} = \!\!\!\!\!\!\sum\limits_{\lfloor m / 2 \rfloor \; \text{integer}} \!\!\!\!\!\!\! \rho_{mm}, \quad p_\text{t}^k = \!\!\!\!\!\!\!\sum\limits_{\lfloor m / 2^k \rfloor \; \text{integer}} \!\!\!\!\!\!\!\!\rho_{mm}, \quad 1 = \sum\limits_m \rho_{mm}.
\end{gather*}
Next, we apply a basis transformation $\mathbf{S}$ on $\rho_\text{D}$ such that $\rho'_\text{D} = \mathbf{S} \cdot  \rho_\text{D}$ with $\rho'_\text{D} = \begin{pmatrix}
		p_\text{q} &
		p_\text{t}^1 &
		\dots &
		p_\text{t}^n & 1 \; \vline \;
		\dots \;\,
	\end{pmatrix}^T$,
leading to the dynamics $\dot{\rho}'_\text{D} = \mathbf{S} \cdot \mathbf{\mathcal{L}}_\text{D} \cdot \mathbf{S}^{-1} \rho'_\text{D}$. Note that the unity in the line above is required to describe the excitations from the ground state. As we prove in \aref{sec_proof_solomon}, on the basis of \eref{eq_inverse_approx_2}, the new rate equation is of the form
\begin{align}
	\mathbf{S} \cdot \mathbf{\mathcal{L}}_\text{D} \cdot \mathbf{S}^{-1} = \left(\begin{array}{ccc}
		\bar{\mathbf{A}} & \hphantom{x} \bar{\mathbf{\Gamma}}_\uparrow & \mathbf{0}\;\cdots\; \mathbf{0}\\[3pt] \hline \\[-5pt]
		\multicolumn{3}{c}{\cdots}  \\[3pt]
	\end{array}\right), \label{eq_coord_transf_rate_eq}
\end{align}
where $\bar{\mathbf{A}}$ is a $(n + 1)$-dimensional square matrix and $\bar{\mathbf{\Gamma}}_\uparrow$ is a $(n + 1)$-dimensional vector. They define the Solomon equations, which are independent of the microscopic structure of the populations, as reflected by the zeros to the right of $\bar{\mathbf{\Gamma}}_\uparrow$. Because of this decoupling, the matrix elements below the horizontal line are irrelevant for the dynamics of the probabilities. $\bar{\mathbf{A}}$ and $\bar{\mathbf{\Gamma}}_\uparrow$ describe the relaxation and excitation processes of the probabilities, respectively. 
The bar on top of $\bar{\mathbf{A}}$ and $\bar{\mathbf{\Gamma}}_\uparrow$ denotes that the rates are altered by the two-photon cross-relaxation rates $\Gamma_\text{qt}^{\sigma_k}$.
The Solomon equations comprise the following new relevant rates: the cross-relaxation rates $\bar{\Gamma}_\text{qt}^k := \Gamma_\text{qt}^{\delta_k} - \Gamma_\text{qt}^{\sigma_k}$, the transition rates of the TLSs
$\bar{\Gamma}^{\text{t}_k}_{\uparrow,\downarrow} := \Gamma_{\uparrow,\downarrow}^{\text{t}_k} + \Gamma_\text{qt}^{\sigma_k}$, and
the qubit transition rates
$\bar{\Gamma}^\text{q}_{\uparrow,\downarrow} := \Gamma_{\uparrow,\downarrow}^\text{q} + \sum_k \Gamma_\text{qt}^{\sigma_k}$.  However, in the usual regime $\Gamma_2^k \ll \omega_\text{q}$ for superconducting qubits, contributions from two-photon processes are negligible.

In our scenario, where the TLS are not interacting with each other, we obtain for $\bar{\mathbf{A}}$ a so-called arrowhead-type matrix so that the Solomon equations are of the form
\begin{gather*}
	\dot{\mathbf{p}}
	= - \!\begin{pmatrix}
		\,\bar{\Gamma}_1^\text{q}  + \sum \bar{\Gamma}_\text{qt}^k \,& -\bar{\Gamma}_\text{qt}^1 & \dots & -\bar{\Gamma}_\text{qt}^k \\[3pt]
		-\bar{\Gamma}_\text{qt}^1 & \bar{\Gamma}_1^1 + \bar{\Gamma}_\text{qt}^1 & & \\[1pt]
		\vdots &&\ddots & \\[5pt]
		- \bar{\Gamma}_\text{qt}^n &&& \bar{\Gamma}_1^n + \bar{\Gamma}_\text{qt}^n
	\end{pmatrix} \! \mathbf{p} + \! \begin{pmatrix}
		\bar{\Gamma}_\uparrow^\text{q}\\[3pt]
		\bar{\Gamma}_\uparrow^1\\[1pt]
		\vdots\\[5pt]
		\bar{\Gamma}_\uparrow^n
	\end{pmatrix}
\end{gather*}
with the usual definition $\bar{\Gamma}_1 = \bar{\Gamma}_{\uparrow} + \bar{\Gamma}_{\downarrow}$.

Finally, we want to note that the proof in \aref{sec_proof_solomon} is also valid when the TLSs are interacting with each other. This means that the Solomon equations are correct as long as the approximations that lead to the cross-relaxation rates in \eref{eq_cross_relaxation} can be justified. In general, it is very difficult to foresee the range of validity when the TLSs become coherent and close in frequency \cite{Muller2009}. At some point a collective behavior of the TLSs comes into play, where the TLSs essentially form a large single spin, as discussed in detail in the context of superradiance \cite{Muller2009, Kessler2010, Shao2023}.

\section{Analytic solutions and approximations} \label{sec_solomon_analytics}
The rich relaxation dynamics of the Solomon equations allow us to draw conclusions on the connectivity and strength of the cross-relaxation rates. In NMR, the measured cross-relaxation rates contain information on the nuclear spins and their distances. This paved the way for the broad field of two-dimensional NMR spectroscopy \cite{friebolin2010}. Similarly, we show in this section that the relaxation dynamics on long timescales contains information on the cross-relaxation distribution. From this distribution, one might be able to draw conclusions on the frequency and spatial distribution of the TLSs and get an idea of their physical nature. For instance, if there is a dipolar coupling between the qubit and the TLSs, one has $g_k \propto 1 / r^3$, with $r$ being the distance to the qubit. When the TLSs are spread on the surface around the qubit we count the number $k(r) \propto r^2$ of TLSs within $r$ and consequently $\Gamma^k_\text{qt} \propto 1 / r_k^6 \propto 1 / k^3$. In case the TLSs are spread in all three dimensions, one obtains $\Gamma^k_\text{qt} \propto 1 / k^2$. For the scenario presented in \fref{fig_Gqt}, when the coupling and the mutual decoherence are approximately constant but the TLSs are spread in frequency, we also expect $\Gamma^k_\text{qt} \propto 1 / k^2$.\\

To simplify the discussion, we assume that the mutual decoherence of the qubit and the TLSs is strong enough ($\Gamma^k_2 > 4g_k$, s. \aref{sec_stochastic_liouville}) for the Solomon equations to be valid.
For superconducting qubits this assumption is not always fulfilled. For instance, it is possible to observe coherent oscillations with dielectric TLSs~\cite{Shalibo2010Oct, Lisenfeld2010Dec}.  For weakly coupled TLSs, e.g. spins, this regime is applicable.
When we compute the cross-relaxation distribution, we can assume that $\Gamma^k_2 \ll \omega_\text{q}$ such that two-photon processes $\Gamma^\sigma_\text{qt}$ can be neglected (\eref{eq_cross_relaxation}). Note that this assumption might not be valid for low-frequency fluxonium qubits \cite{Zhang2021Jan}.
Both assumptions and the form of the cross-relaxation rate \eref{eq_cross_relaxation} imply that the qubit will only exchange energy significantly with those TLSs that are close in frequency, $\delta_k \simeq \Gamma_2^k$.
We will therefore assume that the qubit and the TLSs relax approximately to the same thermal population $p_\text{th} = \Gamma^j_\uparrow / \Gamma^j_1$. Moreover, we assume that all TLSs are of the same physical origin, suggesting a single intrinsic relaxation rate for the TLSs. Under these assumptions, we can now rewrite the Solomon equations: we remove the bar on top of the rates to indicate the neglect of two-photon processes, and for convenience we introduce the intrinsic relaxation rates $\Gamma_\text{q} = \Gamma_1^\text{q}$ and $\Gamma_\text{t} = \Gamma_1^{\text{t}_k}$. The Solomon equations then read
\begin{gather*}
	\dot{\mathbf{p}}
	= - \!\begin{pmatrix}
		\Gamma_\text{q} + \sum \Gamma_\text{qt}^k&\!\! -\Gamma_\text{qt}^1 & \dots &\!\! -\Gamma_\text{qt}^k \\[2pt]
		-\Gamma_\text{qt}^1 &\!\! \Gamma_\text{t}^1 + \Gamma_\text{qt} & & \\
		\vdots &&\ddots & \\[4pt]
		- \Gamma_\text{qt}^n &&&\!\! \Gamma_\text{t}^n + \Gamma_\text{qt}
	\end{pmatrix} \! \mathbf{p} + \! \begin{pmatrix}
		\Gamma_\text{q}\\[2pt]
		\Gamma_\text{t}\\
		\vdots\\[4pt]
		\Gamma_\text{t}
	\end{pmatrix}\!p_\text{th}.
\end{gather*}
Following from \eqsref{eq_def_transition_rates} and \ref{eq_def_relax_rate}, we have for the qubit
\begin{gather}
\Gamma_\uparrow(t) =  \Gamma_\text{q} p_\text{th} + \sum_k  \Gamma_\text{qt}^k p_\text{t}^k(t), \label{eq_up_rate}\\
\Gamma_\downarrow(t) = \Gamma_\text{q}(1 - p_\text{th}) + \sum_k \Gamma_\text{qt}^k (1 - p_\text{t}^k(t)), \label{eq_down_rate} \\
\Gamma_1 = \Gamma_\text{q} + \Gamma_\text{TLSs}, \label{eq_relax_rates}
\end{gather}
where we see once more that the total qubit relaxation rate $\Gamma_1$ is independent of the TLS populations. The qubit relaxation induced by the TLSs is given by $\Gamma_\text{TLSs} = \sum_k \Gamma_\text{qt}^k$. From \eref{eq_def_peq} we obtain
\begin{align}
	p_\text{eq}(t) = \frac{\Gamma_\text{q} p_\text{th} + \sum_k  \Gamma_\text{qt}^k p_\text{t}^k(t) }{\Gamma_1}. \label{eq_equilibrium_pop}
\end{align}
Similarly, we can introduce
\begin{align}
	p_\text{eq}^\text{TLSs}(t) = \frac{\sum_k  \Gamma_\text{qt}^k p_\text{t}^k(t) }{\sum_k \Gamma_\text{qt} ^k},  \label{eq_equilibrium_pop_TLSs}
\end{align}
which would be the equilibrium population of the qubit in the absence of intrinsic qubit loss, and hence it serves as a measure for the TLS population.\\

In the past, numerous techniques have been invented to achieve hyperpolarization, i.e., to create a population that is far from thermal equilibrium. Typically, hyperpolarization is associated with nuclear spins. At first, the nuclear Overhauser effect \cite{Overhauser1953Oct} was predicted and soon after experimentally verified. Polarization was achieved by saturating electronic spins via microwave irradiation, which then cross-relax and polarize the nuclear spins. Soon after, hyperpolarization with optical pumping was discovered. In Ref. \cite{Broadway2018Mar} the authors optically reset a nitrogen spin in a stroboscopic manner to either its ground or excited state, which then cross-relaxes and hyperpolarizes its surrounding nuclear spins.

\begin{figure}[ht]
\begin{center}
\includegraphics[scale=1.0]{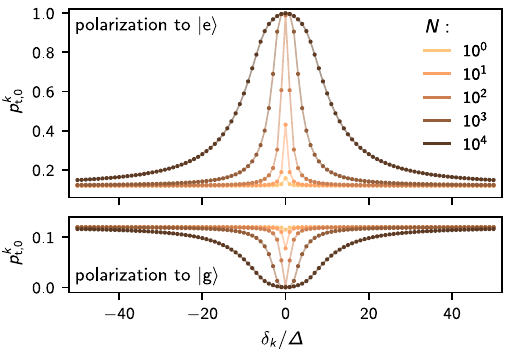}
\caption{\textbf{TLS hyperpolarization}. Initial TLS populations $p_\text{t,0}^k$ plotted versus their detuning $\delta_k$ for various polarization times $N \cdot t_\text{rep}$, with $t_\text{rep}$ being the repetition time of the stroboscopic qubit preparations to $|\text{e}\rangle$ (top panel) and $|\text{g}\rangle$ (bottom panel). The TLS populations were calculated using \eref{eq_TLSstab}.} \label{fig_initial_polarization}
\end{center}
\end{figure}

Making use of design flexibility and fast control of superconducting artificial atoms, hyperpolarization of long-lived TLSs can be achieved in many different ways, for instance by using saturation pulses \cite{Spiecker}. Similarly to Ref.~\cite{Broadway2018Mar}, in the experiments presented later in this section, the TLSs are cooled or heated by stroboscopic qubit preparations using an active feedback protocol, as discussed in detail in our earlier work \cite{Spiecker}.
During this polarization sequence of length $N \cdot t_\text{rep}$, where $t_\text{rep}$ is the repetition time, the qubit population is approximately constant, corresponding to the targeted preparation state. Consequently, the Solomon equations predict an exponential relaxation for the TLSs to their new steady-state value. Immediately after the polarization sequence at $t = 0$, the TLS populations $p_\text{t,0}^k :=  p_\text{t}^k(t = \nolinebreak 0)$ are given by 
\begin{align}
p_\text{t,0}^k = \left(p_\text{th} - p_\text{s}^k\right) e^{-\left(\Gamma_\text{qt}^k + \Gamma_\text{t}\right) N t_\text{rep}} + p_\text{s}^k \label{eq_TLSstab}
\intertext{with steady-state values}
p_\text{s}^k = \frac{\Gamma_\text{t} p_\text{th} + \Gamma_\text{qt}^k}{\Gamma_\text{t} + \Gamma_\text{qt}^k} \qquad \text{or} \qquad p_\text{s}^k = \frac{\Gamma_\text{t} p_\text{th}}{\Gamma_\text{t} + \Gamma_\text{qt}^k} \nonumber
\end{align} 
for heating or cooling of the environment, respectively. These initial TLS populations $p_\text{t,0}^k$ are shown in \fref{fig_initial_polarization} for various polarization times.\\

\begin{figure*}[ht!]
\begin{center}
\includegraphics[scale=1.0]{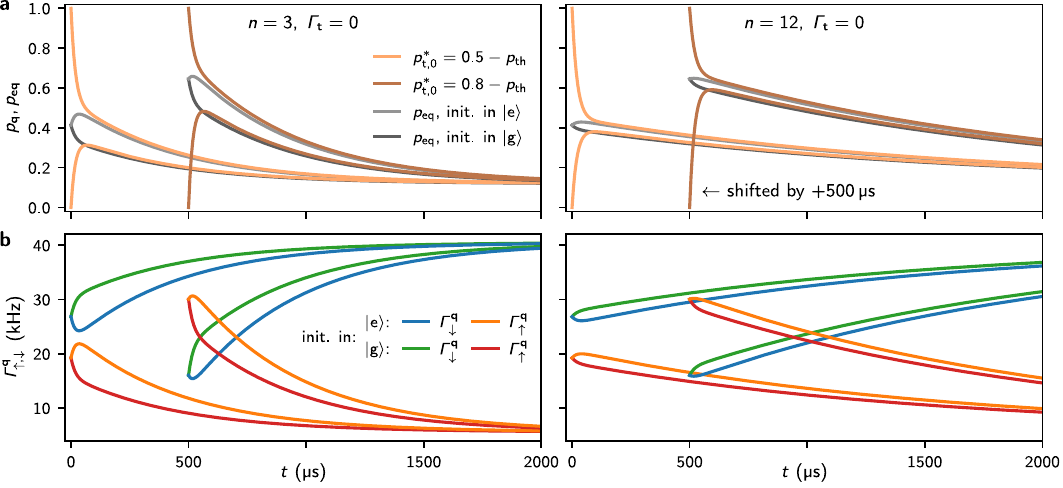}
\caption{\textbf{Qubit relaxation and transition rates for identical cross-relaxation rates to $\mathbf{n}$ TLSs.} \textbf{a} Relaxation dynamics of the qubit population $p_\text{q}$ and equilibrium population $p_\text{eq}$ following qubit initialization to $|\text{g}\rangle$ or $|\text{e}\rangle$. We assume that all $n$ TLSs have the same initial population $p_\text{t,0}^*$. The curves with the higher initial TLS population are shifted in time for better visibility. Between the left and right panel, only the number $n$ is varied. Additionally, we assume the TLSs to be lossless $\Gamma_\text{t} = 0$ and a thermal population $p_\text{th} = 0.12$ for qubit and TLSs. The parameters were chosen to resemble the experimental findings~\cite{Spiecker}.~\textbf{b} Qubit transition rates $\Gamma_{\uparrow, \downarrow}^\text{q}$ corresponding to the relaxation curves shown in panel~a. Not surprisingly, the cooling and heating effect from the qubit initialization, i.e., the difference between, e.g., the orange and red curves, is less pronounced for increasing $n$. All curves are obtained from \eref{eq_relaxation_identical_rates} and using \eqsref{eq_up_rate}, \ref{eq_down_rate} and \ref{eq_equilibrium_pop}.} \label{fig_solomon}
\end{center}
\end{figure*}

The structure of the arrowhead matrix in the Solomon equations entails many useful properties \cite{Shen2009, Jakovcevic2015}. Without loss of generality, we can assume that the rates are nonzero, $\Gamma_\text{qt}^k > 0$, and sorted, $\Gamma_\text{qt}^k \geq \Gamma_\text{qt}^{k+ 1}$. With this definition, we can state one of the most important properties following from the interlacing theorem. It yields immediately the following relations for the eigenvalues:
\begin{gather}
	\lambda_0 > \Gamma_\text{qt}^1 +\Gamma_\text{t} \geq \lambda_1 \geq \dots \geq \Gamma_\text{qt}^n + \Gamma_\text{t} \geq \lambda_{n},  \label{eq_chauchy} \\
	\lambda_0 > \Gamma_1  \geq \lambda_{n}.
\end{gather}
Furthermore, $\lambda_n \geq \min\{\Gamma_\text{q}, \Gamma_\text{t}\}$ is equal if and only if $\Gamma_\text{q} = \Gamma_\text{t}$. The inequality (\eref{eq_chauchy}) shows that the cross-relaxation distribution translates approximately into the same distribution for the eigenvalues of the Solomon equations.
To simplify the following calculations, we introduce the out-of-equilibrium population $\mathbf{p}^* = \mathbf{p} - \mathbf{p}_\text{th}$, where $\mathbf{p}_\text{th} = \begin{pmatrix} p_\text{th}, \dots p_\text{th} \end{pmatrix}^T$ is the steady state for the system in thermal equilibrium.\\

\begin{center}
	\bf \small A. The case of identical cross-relaxation rates
\end{center}
The system of differential equations can be solved analytically for the special case when all cross-relaxation rates are identical, $\Gamma_\text{qt}^k = \Gamma_\text{qt}$. While this case is likely not relevant for dielectric TLSs in superconducting devices, it might be applicable for an environment consisting of hyperfine split spins given their narrow frequency distribution \cite{deGraaf2017Jan}. When the system is driven out of equilibrium by operating the qubit, e.g. following the polarization protocol described earlier, the TLSs will always be populated identically. This means that the $(n - 1)$-fold eigenvalue $\lambda_1 = \Gamma_\text{t} + \Gamma_\text{qt}$ does not take part in the relaxation dynamics, which will therefore be bi-exponential. We derive the analytical solutions in \aref{sec_solomon_solutions}.

\begin{figure}[t!]
\begin{center}
\includegraphics[scale=1.0]{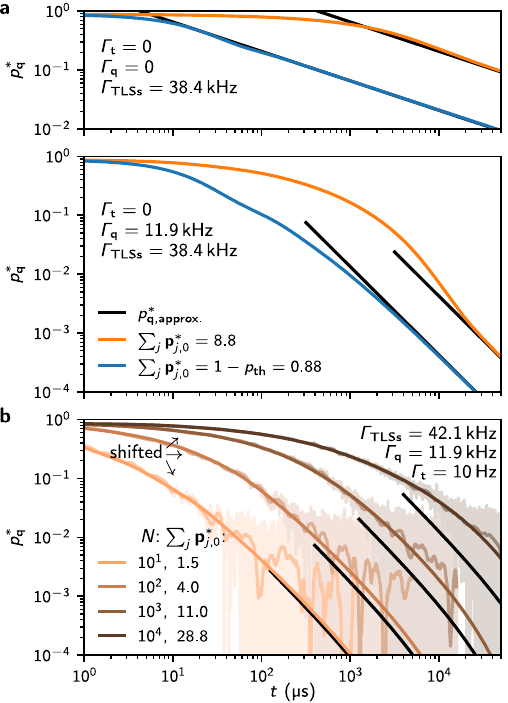}
\vspace*{-1.8em}
\caption{\textbf{Polynomial relaxation.} \textbf{a} Relaxation of the qubit with only the qubit being excited initially (blue lines) or with the first 9 TLSs 
being excited in addition (orange lines). The black curves show the limit solution of either \eqsref{eq_long_term_gamma_zero} or \ref{eq_long_term}. 
We use the distribution $\Gamma_\text{qt}^k = 4ab^2 / k^2$ with $a$ and $b$ as in panel~b.
The top panel shows the scenario without any losses in the system $\Gamma_\text{q} = \Gamma_\text{t} = 0$, simulated with 2000 TLSs, whereas the bottom panel depicts the scenario with the intrinsic qubit loss $\Gamma_\text{q}$ taken from the experiment and simulated with 100 TLSs.
\textbf{b} Qubit relaxation taken from \rcite{Spiecker} for different lengths of the TLS polarization sequence to the excited state. To reduce the noise, a fifth-order Savitzky-Golay filter with increasing window length $w(t) = t / 2$ was applied on the data. The curves $N=10^1$~-~$10^3$ are shifted leftwards for better visibility by factors of $\sqrt{10}$. The continuous lines show the exact result (\eref{eq_general_solution}) using the distribution of \eref{eq_rates_lorenzian_h} with $a = \SI{25.5}{kHz}$, $b = 0.48$ and $c = 0$, slightly updated compared to \rcite{Spiecker}. The initial polarization is modeled with \eref{eq_TLSstab} as shown in \fref{fig_initial_polarization}. The simulation was carried out with 101 TLSs, with more than 50 TLSs required for convergence.
The limit solution \eref{eq_long_term} deviates with increasing initial polarization, since the condition $t \gg 1 / \lambda_n$ with $\mathbf{p}_\text{j,0}^* = 0$ for $j > n$ is not fulfilled (s. text and \fref{fig_initial_polarization}).} \label{fig_loglog}
\end{center}
\end{figure}

The relaxation of the qubit and the corresponding transition rates $\Gamma_\uparrow$ and $\Gamma_\downarrow$ are depicted in \fref{fig_solomon} for different numbers of TLSs with different initial populations.
The qubit initialization, typically to the ground state $p_\text{q,0}^* = 0$ or excited state $p_\text{q,0}^* = 1$, leads to distinct long-term relaxation dynamics. This can be observed in particular for a small number of TLSs, as shown in \frefadd{fig_solomon}{a}.
The qubit initialization does not affect the starting value for the transition rates. However, since the initialization adds or removes one quanta in the system, we notice a difference in the relaxation on longer timescales, as shown in \frefadd{fig_solomon}{b}.
The bi-exponential relaxation in this scenario is qualitatively similar to the experimental results presented in \rcite{Spiecker}, but insufficient to describe the slow and non-exponential relaxation observed on long timescales (cf. \frefadd{fig_loglog}{b}). These effects can only be explained when the rates $\Gamma_\text{qt}^k$ are not identical. Instead, a few fast and many slow cross-relaxations rates are required. 

We want to add that in the limit of infinitely many weakly coupled TLSs such that $\Gamma_\text{TLSs}$ is constant, we have $\lambda_{0, 2} \in \{ \Gamma_1, \Gamma_\text{t} \}$. Then it follows from \eref{eq_relaxation_identical_rates} that the TLSs evolve independently of the qubit with $p^*_\text{t}(t) = p^*_\text{t,0} e^{-\Gamma_\text{t} t}$, while the qubit dynamics remains bi-exponential, governed by
\begin{align}
\dot{p}_\text{q} = - \Gamma_1 p_\text{q} + \Gamma_\text{TLSs} p_\text{t,0}^* e^{-\Gamma_\text{t} t} + \Gamma_1 p_\text{th},
\end{align}
as may also be read from the Solomon equations. 
To be more precise, in the interesting scenario in which the TLSs are long-lived, $\Gamma_\text{t} < \Gamma_1$, we have $\lambda_2 = \Gamma_\text{t} + \Gamma_\text{TLSs} (\Gamma_\text{q} - \Gamma_\text{t}) / (\Gamma_1 - \Gamma_\text{t}) / n + \mathcal{O}(1 / n^2)$ and for the TLSs it then holds that $p^*_\text{t}(t) = p^*_\text{t,0} e^{- \lambda_2 t} + \mathcal{O}(1 / n)$. In the limit $n \gg 1$ with the TLSs being initially in thermal equilibrium $p^*_\text{t,0} = 0$, their relaxation thus becomes asymptotically decoupled from the qubit decay, justifying the Born approximation in this limit.\\

\begin{center}
	\bf \small B. The case of distributed cross-relaxation rates
\end{center}
Without loss of generality, we will assume that all the transfer rates are non-zero and distinct from each other. This is justified because if relaxation rates happen to be identical, their eigenvalues do not individually take part in the relaxation dynamics of the qubit and can be collapsed with Givens rotations, as illustrated in the previous case analysis. The eigenvalues $\lambda_m$ of an irreducible arrowhead matrix are given as the roots of the so-called Pick function \cite{Jakovcevic2015}:
\begin{align}
f(\lambda) = \Gamma_\text{q} + \sum \Gamma_\text{qt}^k - \lambda - \sum \frac{{(\Gamma_\text{qt}^k})^2}{ \Gamma_\text{qt}^k + \Gamma_\text{t} - \lambda} = 0. \label{eq_pick}
\end{align}
The corresponding eigenvectors can be expressed via
\begin{multline}
v_m = \big(1 \quad \Gamma_\text{qt}^1 / (\Gamma_\text{qt}^1 + \Gamma_\text{t} - \lambda_m) \quad \dots \\ \dots \quad \Gamma_\text{qt}^n /(\Gamma_\text{qt}^n + \Gamma_\text{t} - \lambda_m) \big)^T. \label{eq_vec}
\end{multline}
Fortunately, one can find that $\Vert v_m \Vert^2 = - \frac{\partial f}{\partial \lambda} \Big\vert_{\lambda_m}$.
The fundamental solution of the Solomon equations from \sref{sec_solomon_analytics} can now be written as
\begin{align}
\mathbf{p}(t) =& \begin{pmatrix}
1 / \Vert v_0 \Vert & \dots & 1 / \Vert v_n \Vert  \\
\vdots &  & \vdots & \\
 && \\
\end{pmatrix}\cdot 
 \begin{pmatrix}
e^{- \lambda_0 t} & & \\
 & \ddots &  & \\
 && \\
\end{pmatrix} \nonumber \\
&\;\cdot \begin{pmatrix}
1 / \Vert v_0 \Vert & \dots &  \\
\vdots &  &  & \\
1 / \Vert v_n \Vert & \dots& \\
\end{pmatrix} \mathbf{p}^*_0 + \mathbf{p}_\text{th}. \label{eq_general_solution}
\end{align}
When initially only the qubit is out of equilibrium, we have
\begin{align}
p_\text{q}(t) = \sum\limits_{m = 0}^n \frac{e^{- \lambda_m t}}{\Vert v_m \Vert^2} \,p_{\text{q},0}^* + p_\text{th}  \label{eq_pq_analytical}
\end{align}
and from \eref{eq_equilibrium_pop} with the help of \eref{eq_pick} we find
\begin{align}
p_\text{eq}(t) &= \sum\limits_{m = 0}^n \frac{\Gamma_1 - \lambda_m}{\Gamma_1} \cdot \frac{e^{- \lambda_m t}}{\Vert v_m \Vert^2} \,p_{\text{q},0}^* + p_\text{th},  \label{eq_peq_analytical}
\end{align}
where one may recognize \eref{eq_relax_qubit} when inserting \eref{eq_pq_analytical} in \eref{eq_peq_analytical}.

\begin{center}
	\bf \small C. Polynomial relaxation
\end{center}
In the following, we will discuss possible long-term relaxation behaviors of the system.  Since $p_\text{q}$ and $p_\text{eq}$ decay in a similar way on long timescales (s. \eqsref{eq_pq_analytical} and \ref{eq_peq_analytical}) we will use $p_\text{q}(t)$ to discuss the relaxation, which is 
directly accessible in experiments.
In particular, we are interested in the emergence of non-exponential relaxation curves that must arise by virtue of \eref{eq_pq_analytical} from the sum of many exponential functions. Obviously, if the qubit interacts with a finite number of TLSs, the relaxation can be approximated on long timescales by $p_\text{q}(t) \approx \alpha e^{-\lambda_n t}$ and the non-exponential behavior can only appear for $t < 1 / \lambda_n$. An exponential relaxation will also be seen for an infinite number of TLSs that have a finite lifetime $\Gamma_\text{t} > 0$, in which case we obtain $p_\text{q}(t) \approx \alpha  e^{- \Gamma_\text{t} t}$ on long timescales, as can be deduced from \eref{eq_chauchy}, and therefore a non-exponential behavior can only appear for $t < 1 / \Gamma_\text{t}$. 
Thus, a non-exponential relaxation appears on long timescales for a large number of TLSs that are long-lived, $\Gamma_\text{t} \approx 0$, and with eigenvalues $\lambda_i$ that vanish continuously with $\lambda_i - \Gamma_\text{t} \rightarrow 0$.

As an insightful example, we will now discuss the experimental situation in which the TLSs are spread in frequency. For the modeling, we will assume that the TLSs are equally spaced in frequency with spacing $\Delta$. In this case, the cross-relaxation rates (\eref{eq_cross_relaxation}) are given by 
\begin{align}
\Gamma_\text{qt}^h = \frac{a b^2}{b^2 + (h -  b c)^2} \quad \text{with} \quad h \in \mathbb{Z}, \label{eq_rates_lorenzian_h}
\end{align}
where $a = 2 g^2 / \Gamma_2$, $b = \Gamma_2 / \Delta$, $c = \Delta_0/ \Gamma_2$ and $\Delta_0$ describes a frequency shift of the TLSs with respect to the qubit frequency. Due to the periodicity, $\Delta_0$ can be restricted to $\Delta_0 \in \{0, \Delta / 2\}$. We use the index $h$ to clarify that these rates are not sorted (\eref{eq_chauchy}).
In case $b \gg 1$, which corresponds to a high TLS density, many TLSs interact with the qubit with a similar rate. Here, a good approximation on moderate timescales is given by case study A.
In the case $b \lesssim 1$, as in our experiment, we have the interesting situation in which the rates vanish continuously.

Next, we compute the relaxation dynamics on long timescales following from \eref{eq_pq_analytical}. For the distribution in \eref{eq_rates_lorenzian_h} the analysis below can only be applied for the special cases $\Delta_0~=~0$ (one TLS in resonance with the qubit), $\Delta_0~=~\Delta/4$ (equally spaced detunings $|\delta_k|$), and $\Delta_0~=~\Delta / 2$ (maximum detuning between the closest TLSs and the qubit). For more details, see \aref{sec_special_solutions}. However, as we are mainly interested in the long-time dynamics, we can slightly approximate the rate distribution to yield a simpler and more instructive analysis.
We use
\begin{gather}
\Gamma_\text{qt}^k = a / k^2 \quad \text{with} \quad k \in \mathbb{N}^+  \label{eq_rates_quadratic} \\
\rightarrow \quad \Gamma_\text{TLSs} = \sum \Gamma_\text{qt}^k  = \frac{a \pi^2}{6}. \nonumber
\end{gather}
Note that for a direct comparison with the distribution defined in \eref{eq_rates_lorenzian_h} the parameter $a$ needs to be scaled by a factor of 4. With the new distribution, the Pick function can be simplified and expressed in a closed form:
\begin{align}
f(\lambda) &=\Gamma_\text{q} - \Gamma_\text{t} - \frac{\lambda'}{2} - \frac{\pi a}{2}\sqrt{\frac{\lambda'}{a}} \cot \sqrt{\frac{\pi^2 a}{\lambda'}} \label{eq_pick_cot}
\end{align}
with $\lambda' = \lambda - \Gamma_\text{t}$. Furthermore, we can compute $\Vert v_m \Vert^2$ using the fact that by definition $f(\lambda_m) = 0$, which yields
\begin{align}
\Vert v_m \Vert^2 &= - f'(\lambda){\Big\vert}_{\lambda_m} = - \frac{\partial f}{\partial \lambda'} {\Big\vert}_{\lambda'_m} \nonumber \\
                  &= \frac{1}{2} + \frac{3 \Gamma_\text{TLSs} - (\Gamma_\text{q} - \Gamma_\text{t})}{2 a} z_m + \frac{(\Gamma_\text{q} - \Gamma_\text{t})^2}{a^2}z_m^2 \nonumber \\
                  &= 1/2 + \beta z_m + \gamma^2 z_m^2. \label{eq_v2}
\end{align}
Here, we further introduced $z_m = a / \lambda_m'$. We will see that the linear and quadratic terms in $z_m$ give rise to different long-term relaxation dynamics.

The next step is to evaluate the sum in \eref{eq_pq_analytical} on long timescales. The derivation is presented in \aref{sec_solomon_solutions}. In the situation in which $\gamma^2 = 0$, which essentially describes the dilution of the initial qubit excitation into the TLS environment, the long-term relaxation dynamics is governed by
\begin{align}
p_\text{q,approx.}^*(t) = \lim_{t \rightarrow \infty} \; p_\text{q}^*(t) = \frac{\sqrt{\pi}}{2 \beta}\frac{e^{- \Gamma_\text{t} t}}{(at)^{1 / 2}} \;p_{\text{q},0}^*.  \label{eq_long_term_gamma_zero}
\end{align}
In the experimentally more likely situation in which $\gamma^2~>~0$, 
the relaxation dynamics becomes
\begin{align}
p_\text{q,approx.}^*(t) = \lim_{t \rightarrow \infty} \; p_\text{q}^*(t) = \frac{\sqrt{\pi}}{4 \gamma^2}\frac{e^{- \Gamma_\text{t} t}}{(at)^{3 / 2}} \;p_{\text{q},0}^*.  \label{eq_long_term}
\end{align}
The long-time relaxation dynamics is likely to be hidden in the noise when initially only the qubit is brought out of equilibrium. A simple way to improve the visibility is to also initialize a few of the most resonant TLSs, as in the experiment (s. \fref{fig_initial_polarization}).
Then, for a large number of TLSs where we have $\lambda'_m \rightarrow 0$, one can deduce from \eqsref{eq_vec} and \ref{eq_general_solution} that the long-time solution is still valid. 
We just have to replace $p^*_{\text{q},0}$ with the total out-of-equilibrium excitation $\sum_j \mathbf{p}^*_{j,0}$.
The long-term solution sets in under the condition $t \gg 1 / \lambda_n$ with $n$ such that $\mathbf{p}^{*}_{j, 0} = 0$ for $j > n$.
In \frefadd{fig_loglog}{a} we show several relaxation curves verifying the power-law decay on long timescales, which can also be observed in the experiment (s. \frefadd{fig_loglog}{b}). However, in the experiment the above condition is only approximately fulfilled, since all TLSs were at least partially excited during the polarization sequence (s. \fref{fig_initial_polarization}).\\

\begin{center}
	\bf \small D. Generalization
\end{center}
As discussed in the beginning of this section, one can expect also other cross-relaxation distributions in experiments. 
In the following, we will therefore discuss the long-time behavior for a general cross-relaxation distribution of the form
\begin{gather}
	\Gamma^k_\text{qt} = a / k^d \quad  \text{with $d > 1$ and $k \in \mathbb{N}^+$} \label{eq_gamma_d} \\
	\rightarrow \quad \Gamma_\text{TLSs} = \sum \Gamma_\text{qt}^k  = \zeta(d), \nonumber
\end{gather}
with $\zeta$ being the Riemann zeta function. The derivation of the limit behavior is similar to the one presented above and is given in \aref{sec_solomon_solutions}. The difficulty is to determine the value of $\beta$, for which analytical expressions can be derived for integer values of $d$, as shown in \aref{sec_special_solutions}. For $\gamma = 0$, the relaxation dynamics approaches
\begin{align*}
p_\text{q,approx.}^*(t) = \lim_{t \rightarrow \infty} \; p_\text{q}^*(t) = \frac{\Gamma(1 + \frac{1}{d})}{\beta}\frac{e^{- \Gamma_\text{t} t}}{(at)^{1 / d}} \;p_{\text{q},0}^*.
\end{align*}
Here, $\Gamma$ denotes the Gamma function. For $\gamma^2 > 0$ it becomes
\begin{align*}
p_\text{q,approx.}^*(t) = \lim_{t \rightarrow \infty} \; p_\text{q}^*(t) = \frac{\Gamma(2 - \frac{1}{d})}{\gamma^2 d}\frac{e^{- \Gamma_\text{t} t}}{(at)^{2 - 1 / d}} \;p_{\text{q},0}^*.
\end{align*}
For $d \gg 1$, this solution sets in very late and will be difficult to observe in the experiment, unless the system is approximately lossless $\gamma \rightarrow 0$.

\section{Qubit relaxation as a function of the mutual decoherence} \label{subs:T1_one_TLS}

\begin{figure}[t]
	\begin{center}
		\includegraphics[scale=1.0]{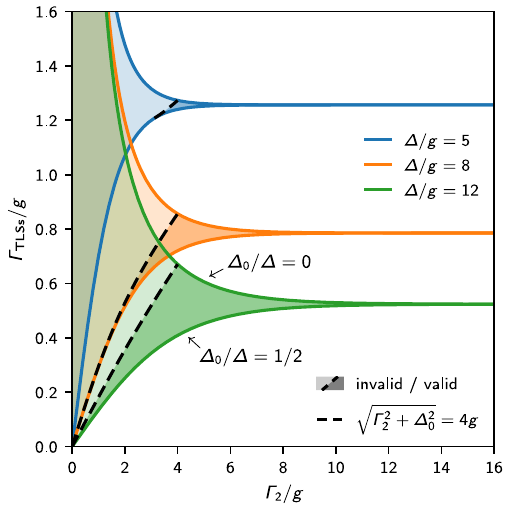}
		\caption{\textbf{TLS-induced qubit relaxation.} $\Gamma_\text{TLSs}$ as a function of the mutual decoherence $\Gamma_2$ for various TLS frequency densities $1 / \Delta$. The shift of the TLS ladder with respect to the qubit frequency is encoded in $\Delta_0$, i.e., the detuning to the most resonant TLS. The dashed line indicates the threshold above which the calculation of the cross-relaxation rate \eref{eq_gamma_qt} is no longer valid. In this region, the qubit and the most resonant TLSs undergo damped vacuum Rabi oscillations (s. \aref{sec_stochastic_liouville} for more details).} \label{fig_gamma_1}
	\end{center}
\end{figure}

In this section, we discuss the qubit relaxation as a function of the mutual decoherence $\Gamma_2$.
From \eref{eq_cross_relaxation} we see that the added qubit relaxation induced by a single TLS is a non-monotonic function in $\Gamma_2$, which vanishes in both limits of $\Gamma_2 \rightarrow 0$ and $\Gamma_2 \rightarrow \infty$. 
The contribution to the relaxation is maximized when $\Gamma_2 = \delta$, amounting to $ \Gamma_\text{qt} = g^2 / \delta$. When the TLSs are spread in frequency with a single coupling strength $g$ and mutual decoherence rate $\Gamma_2$, the sum of the distribution in \eref{eq_rates_lorenzian_h} can be evaluated analytically and results in
\begin{align}
\Gamma_\text{TLSs} &= \!\!\sum_{h = - \infty}^\infty \frac{a b^2}{b^2 + (h - b c)^2} \nonumber \\
&= \pi a b \, \frac{\sinh(2 \pi b)}{\cosh(2 \pi b) - \cos(2\pi b c)} \label{eq_gamma_tlss_final}
\end{align}
with $a, b$ and $c$ as defined in \eref{eq_rates_lorenzian_h}.
Since $\Gamma_2$ cancels in the prefactor $ab = 2 g^2 / \Delta$,
for sufficient decoherence such that $b~\gtrsim~1$, we recover the result of the Bloch-Redfield master equation:
\begin{align*}
\Gamma_\text{TLSs} &\approx \frac{2 \pi g^2}{\Delta} = 2\pi g^2\nu_\text{t},
\end{align*}
with $\nu_\text{t}$ being the TLS density. In \fref{fig_gamma_1} we plot the qubit relaxation as a function of the mutual decoherence for various TLS densities. Note, if the qubit is mainly coupled to a single TLS, the qubit relaxation can be improved by increasing the decoherence of the TLS, e.g., by changing the ambient temperature or applying saturation pulses on the TLS. However, this improvement is lost for multiple TLSs, since the spectral broadening due to the increased $\Gamma_2$ exposes the qubit to far detuned TLSs.
Another method for increasing $\Gamma_2$ and changing qubit relaxation is to exploit the photon shot noise dephasing during qubit readout, as recently demonstrated in \rcite{thorbeck2023}. 

\section{Quantum Jump Statistics} \label{sec_quantum_jumps}

\begin{figure}[t!]
\begin{center}
\includegraphics[scale=1.0]{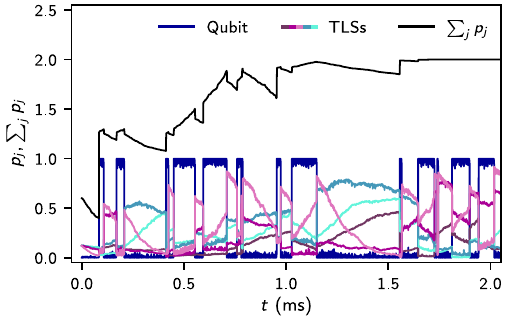}
\caption{\textbf{Illustration of the measurement backaction.} Simulation of a stroboscopic quantum jump trace using the  stochastic Schrödinger equation. The qubit is measured with a repetition time $t_\text{rep} = \SI{2}{\micro s}$. In the simulation shown here, the qubit and the five most resonant TLSs are assumed to be lossless but exposed to stochastic dephasing. At $t = 0$ the system is in a product state with populations $p_j = p_\text{th}$ and random phases. Eventually, the system will be trapped, here in the two-photon manifold, where the total population (black lines) remains constant. We use the parameters of the experiment (\frefadd{fig_loglog}{b}) and the dephasing rates $\Gamma_\varphi^\text{q} = \SI{0.5}{MHz}$ and $\Gamma_\varphi^{\text{t}_k} = \SI{1.0}{MHz}$.} \label{fig_backaction}
\end{center}
\end{figure}

The quantum jumps of a qubit in a Born-Markovian environment are Poisson-distributed, i.e., the qubit undergoes an exponential relaxation. 
When the qubit is coupled to a finite-size TLS environment, a non-Poissonian statistics can be expected due to measurement-induced temperature fluctuations in the TLS environment (\fref{fig_backaction} and \rcite{Moreira2023Dec}), which we visualize by taking a histogram of the qubit dwell times in the excited or ground state (\frefadd{fig_intro}{b}). 
The quantum jump statistics are extracted from quantum jump traces obtained by stroboscopic projective qubit measurements at equidistant intervals. We used the same fluxonium device described in \rcite{Spiecker}. 

\begin{figure}[t!]
\begin{center}
\includegraphics[scale=1.0]{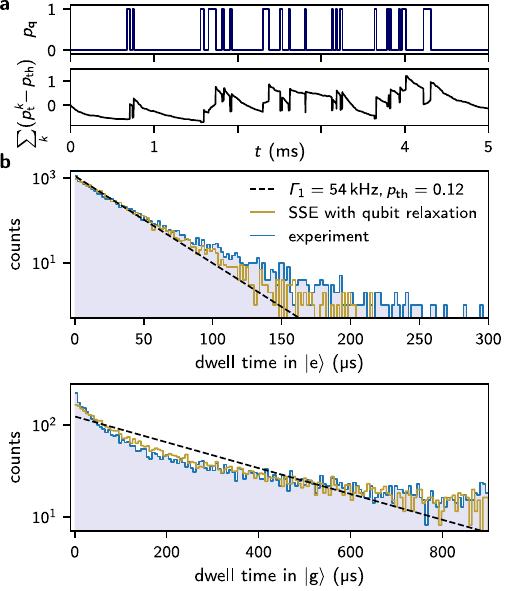}
\caption{\textbf{Comparison between SSE-simulated and measured quantum jumps.} \textbf{a} SSE simulation of a quantum jump trace similar to \fref{fig_backaction}, but including qubit relaxation and the seven most resonant TLSs. The cross-relaxation of the remaining TLSs was added to the intrinsic qubit loss. \textbf{b}~Measured and SSE-simulated quantum jump statistics of the qubit dwell times in $|\text{e}\rangle$ and $|\text{g}\rangle$, respectively. The histograms are generated as described in \fref{fig_intro}. The black dashed line shows the exponential distribution for a Born-Markovian qubit environment with qubit relaxation $\Gamma_1 = \Gamma_\text{q} + \Gamma_\text{TLSs}$ in thermal equilibrium corresponding to $p_\text{th}$.}
\label{fig_SSE_statistics}
\end{center}
\end{figure}

\begin{figure*}[t!]
\begin{center}
\includegraphics[scale=1.0]{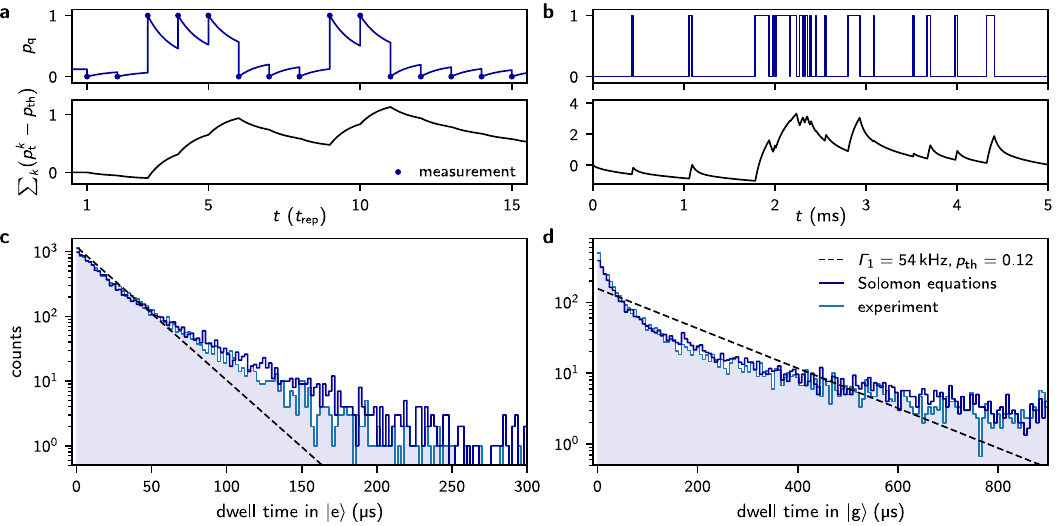}
\caption{\textbf{Comparison between Solomon-equations-simulated and measured quantum jumps.} \textbf{a} Schematic simulation of a quantum jump trace using the Solomon equations. The blue dots show the positions and outcomes of the measurement at $t_\text{rep}$ intervals. The qubit is projected to its eigenstates with the probability corresponding to the qubit population $p_\text{q}$. In the lower panel, we show the excess population of the TLSs, assuming that the TLS populations remain unchanged by the qubit measurement (s. main text for the discussion of the measurement backaction). In between the measurements, the system evolves according to the Solomon equations (\eref{eq_general_solution}).
\textbf{b} Simulation of a longer quantum jump trace following the same method as in panel~a with $t_\text{rep} = \SI{2}{\micro s}$ and using the same parameters as in \frefadd{fig_loglog}{b}. Here, we only show the qubit measurement outcome and the excess TLS population at the time of the measurement. Due to the heat capacity of the environment and the relatively low thermal population $p_\text{th} = 0.12$, the quantum jumps appear bunched when the TLS temperature is increased.
\textbf{c}, \textbf{d} Measured and Solomon-equations-simulated quantum jump statistics of the qubit dwell times in $|\text{e}\rangle$ and $|\text{g}\rangle$, respectively. 
The histograms are generated as described in \fref{fig_intro}. The black dashed line shows the exponential distribution for a Born-Markovian qubit environment. The experimental data presented here plus additional quantum jump traces are publicly available \cite{Spiecker2023Jul}.} \label{fig_quantum_jumps}
\end{center}
\end{figure*}

In this section, we show that the measured quantum jump statistics can be reproduced using a diffusive stochastic Schrödinger equation (SSE) (\fref{fig_SSE_statistics}) and also, somewhat surprisingly, using the Solomon equations (\fref{fig_quantum_jumps}). While the SSE offers a more complete picture by tracking the entanglement of the qubit and the TLSs, the stochastic evolution of the wave function becomes computationally demanding with increasing system size. In addition, the SSE is not unique as it depends on the system details. In a nutshell, the quantum-mechanical challenge is to describe the flow of excitation and energy during the measurement process, in particular when additional environments of the qubit \cite{Stevens2022Sep} and the TLSs are included.

Using the Solomon equations implicitly assumes that the measurement backaction on the TLSs does not change their population expectation values. 
Therefore, one cannot expect an agreement with the SSE in general.
Indeed, we show numerically in \aref{sec_SSE} and experimentally in \rcite{Spiecker2024} that the Solomon equations can lead to distinct quantum jump distributions for a small number of TLSs that dominate the intrinsic qubit loss. This allows us to differentiate between a quantum and classical behavior. Measuring the quantum jump statistics can therefore be used to identify a reduced measurement backaction as well as to investigate a quantum-to-classical transition with an increasing number of TLSs.

Before proceeding with the experimental results, we will clarify in the following the role of the measurement backaction.
At this point it should be emphasized that the TLSs in the experiment, while being approximately lossless, provide the main loss mechanism for the qubit. For now, we can therefore neglect additional environments and consider the closed qubit-TLS system. Let $\ket{\psi_\text{i}}$ and $\ket{\psi_\text{f}}$ denote the wave functions before and after a projective qubit measurement, respectively.
Following a textbook quantum-mechanical measurement on the qubit, $\ket{\psi_\text{f}}$ will be a product state with the qubit being projected either to its ground state $\ket{\text{g}}$ or to its excited state $\ket{\text{e}}$. We have
\begin{align*}
  \ket{\psi_\text{f}} = \left\{ 
  \begin{array}{ll}
       \ket{\text{e}} \otimes \ket{\psi_\text{TLSs}'}  & \text{with prob. } p_\text{q}, \\
       \ket{\text{g}} \otimes \ket{\psi_\text{TLSs}''}  & \text{with prob. } 1 - p_\text{q}, 
  \end{array} \right.
\end{align*}
with the TLS wave function altered depending on the measurement outcome. In this setting, we highlight three scenarios representing very different backaction properties.

In the first scenario, we show that the excitation difference of the qubit before and after the measurement can be provided by the TLSs. Let $\mathcal{M}(m)$ denote the $m^\text{th}$ excitation manifold. 
Then, if $\ket{\psi_\text{i}} \in \mathcal{M}(m)$, it follows that $\ket{\psi_\text{f}} \in \mathcal{M}(m)$ and the excitation number is conserved. 
Here, the measurement process does not change the excitation number, but if the qubit and TLS photon energies are different, it must account for the energy difference.

The second scenario is essentially opposite to the first and contains those wave functions where the expectation values of the TLS populations remain unchanged by the qubit measurement. For example, if the qubit and the TLSs are in the non-entangled product state, then it holds
\begin{align*} 
  \ket{\psi_\text{i}} &= \left(\sqrt{1 - p_\text{q}}\ket{\text{g}}  + \sqrt{p_\text{q}}e^{i \varphi} \ket{\text{e}}\right) \otimes \ket{\psi_\text{TLSs}} \\
 \rightarrow \quad \ket{\psi_\text{f}} &= \left\{ 
  \begin{array}{ll}
       \ket{\text{e}} \otimes \ket{\psi_\text{TLSs}}  & \text{with prob. } p_\text{q} \\
       \ket{\text{g}} \otimes\ket{\psi_\text{TLSs}} & \text{with prob. } 1 - p_\text{q}. 
  \end{array} \right.
\end{align*}
Here, the excitation difference of the qubit must be provided entirely by the measurement process.

Lastly, in general, in the third scenario the excitations can change arbitrarily. Consider, for instance, the following wave function, which is a mixture of two excitation manifolds:
\begin{align*}
\ket{\psi_\text{i}} = \sqrt{1 - p_\text{q}} \ket{\text{g}} \otimes  \ket{\varphi} + \sqrt{p_\text{q}} e^{i \varphi} \ket{\text{e}} \otimes  \ket{\chi},
\end{align*}
where $\ket{\varphi} \in \mathcal{M}(m)$ and $\ket{\chi} \in \mathcal{M}(n)$.  After the measurement, the final wave function will be either in $\ket{\psi_\text{f}} \in \mathcal{M}(m)$ or in $\ket{\psi_\text{f}} \in \mathcal{M}(n + 1)$. 
Here, we see that the total excitation difference must be provided by the measurement process, and 
that repeated measurements on the qubit have a tendency to steer the system into an excitation manifold. For several TLSs in higher excitation manifolds, this purification will become less effective and will likely not compete with relaxation processes that provide transitions between neighboring manifolds. 
For an experimental verification of a spin bath purification, see \rcite{Dasari2022Dec}.

To clarify the previous discussion about the measurement backaction, we depict in \fref{fig_backaction} a stroboscopic quantum jump trace simulated with the diffusive SSE. Since in the discussion the system was considered to be closed, we only allow for dephasive processes (s.~\aref{sec_SSE} for more details). To illustrate the different backaction scenarios, the system wave function was initialized  in a product state with populations $p_j = p_\text{th}$ and random phases.
The fluctuating and varying excitation in the system comes solely from the measurement process indicating mainly scenario three, but also scenario two in the very beginning. Eventually, these fluctuations come to an end when the system state is trapped in one of the photon manifolds; in the case shown here, the system state is trapped in the two-photon manifold. From here on, the dynamics are described by the first scenario.

Due to the trapping behavior, the lossless SSE is not suited to simulate longer quantum jump traces with a specific average qubit population corresponding to the temperature.
In reality, this temperature is defined by additional qubit environments that are responsible for its intrinsic relaxation, essentially providing transitions between excitation manifolds of the qubit-TLS system.
We show in \aref{sec_SSE} various lossless SSE-simulated quantum jump statistics for an even larger system consisting of 16 TLSs trapped in the three-photon manifold and therefore $p_\text{th} = 3 / 16$. 
In some of these simulations, we use the majority of the TLSs to emulate the intrinsic qubit relaxation. Alternatively, we find that the same statistics can be obtained much faster by using a diffusive SSE \cite{Gisin1992Nov, Castin1993Jun, breuer2002theory} that incorporates the qubit relaxation (\aref{sec_SSE}). 
We show in \fref{fig_SSE_statistics} the result of such a diffusive SSE, which reproduces the measured quantum jump statistics (s. \aref{sec_SSE} for more details). The computational complexity could be kept on an acceptable level by recognizing that in particular the most resonant TLS contributes dominantly to the non-Poissonian quantum jump statistics. Accordingly, the simulation was performed for the qubit and the seven most resonant TLSs. 
As a perspective, it would be enlightening to examine other master equation unravelings of the qubit-TLS system of increasing size, while taking into account additional environments if needed. In this context, one may also utilize recent developments for the efficient simulation of larger systems~\cite{Chruscinski2013Jul, Pucci2016May, Zhu2019Aug, Link2023Sep, Xu2023Jul}.

In case the second backaction scenario is realized, the Solomon equations can be used to accurately explain the measured quantum quantum jump statistics.
Suppose the measurement process is not changing the TLS populations. Then, we can integrate the Solomon equations to obtain the qubit population at the time of the upcoming measurement and obtain the probabilities of the measurement outcome.
Starting in thermal equilibrium, we can generate a quantum jump trace and compute the fluctuating energy in the TLS bath that in turn influences the measurement outcome of the qubit (s. \frefadd{fig_quantum_jumps}{a, b}).
The resulting martingales for the energies in the system are similar to Pólya's urn model, except that the qubit and the TLSs can (i) saturate and (ii) decay into their environments. 
The quantum jump statistics generated in this way is contrasted in \frefadd{fig_quantum_jumps}{c,~d} with the experiment and shows quantitatively an even better agreement than the SSE in \frefadd{fig_quantum_jumps_SSE}. 
Importantly, we want to note that measuring non-Poissonian quantum jump statistics hampers the accurate extraction of the qubit's transition rates~$\Gamma_{\uparrow,\downarrow}$ (s. \frefadd{fig_quantum_jumps}{c, d} and \frefadd{fig_quantum_jumps_SSE}{a}).

\section{Conclusion} \label{subs:conclsion}
We presented a thorough derivation of the qubit dynamics in contact with a two-level system environment. An infinite number of weakly coupled TLSs provides a Born-Markovian environment for the qubit. In this case, the well-known Bloch-Redfield master equation is the method of choice to describe relaxation and decoherence of the qubit. The situation changes drastically for a finite number of long-lived TLSs, as the Born approximation is no longer valid. For this case we presented a detailed derivation of the Solomon equations, applicable in the absence of coherent interactions. In contrast to the Bloch-Redfield master equation, the Solomon equations do not require the rotating-wave or secular approximation and they are therefore also valid for low-frequency qubits~\cite{Zhang2021Jan}.

The accuracy of the Solomon equations is demonstrated by the measured and computed qubit relaxation on timescales more than three orders of magnitude longer than the qubit lifetime. We calculated and experimentally confirmed the power-law relaxation behavior on long timescales.
The rich relaxation dynamics contains information on the cross-relaxation distribution. This allows drawing conclusions on the frequency and spatial distribution of the TLSs and can provide insights into their physical nature. In a last step, we showed that both the stochastic Schrödinger equation and the Solomon equations can predict the measured non-Poissonian quantum jump statistics. This finding expands the common expectation that non-Poissonian statistics stems from fluctuations of the qubit’s relaxation time, e.g., a fluctuating number of quasiparticles in the superconductor \cite{Vool2014Dec}. 

Our derivation of the Solomon equations opens new avenues to compute higher-order cross-relaxation pathways that arise from the finite lifetime of the TLSs and from interactions between the TLSs~\cite{Kowalewski2017Dec}. 
In future work, it would certainly be interesting to expand the analysis to describe the qubit's decoherence and frequency shift induced by the discrete TLS environment.
Extending the derivation for more complex interactions and environments such as spins with larger quantum number requires additional approximations and the consideration of ancillary expectation values in the equation of motion~\cite{Kumar2000Sep}. 
A general yet formal description of relaxation, including a discussion on irradiation, can be found in \rcite{Goldman2001Apr}. Furthermore, the Solomon equations suggest the applicability of the nuclear Overhauser effect spectroscopy method on quantum processors when the qubits are tuned in resonance. In this way, one can identify spurious couplings between the qubits.

Last but not least, we believe that tracking the energy flow in multipartite systems~\cite{Karimi2020Apr, Pekola2022Feb} is a powerful tool to understand the backaction of quantum measurements. Analyzing the quantum jump statistics as presented allows us to investigate a quantum-to-classical transition with increasing number of TLSs. Anomalies in the measurement backaction, which could for instance be the result of an accelerated thermalization, have far-reaching consequences because uncorrelated measurement backaction is a key ingredient for quantum error-correction.

\section*{Acknowledgements}
We thank Mitchell Field for a critical reading of the manuscript and for his suggestions for improvement.
This work was supported by the German Ministry of Education and Research (BMBF) within the project QSolid (FKZ: 13N16151). A.S. was supported by the Baden-Württemberg-Stiftung (Project QuMaS).

\appendix
\renewcommand{\thefigure}{\thesection\arabic{figure}}

\section{Liouvillian in superoperator notation} \label{sec_liouvillian}

\begin{figure}[b!]
\begin{center}
\includegraphics[scale=1.0]{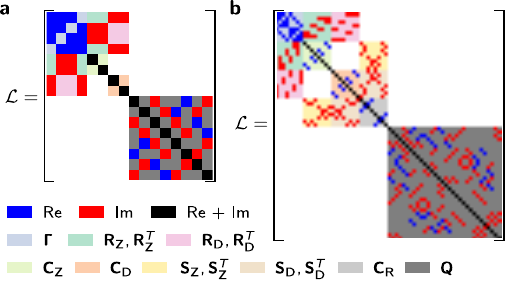}
\caption{\textbf{Liouvillian matrix structure.} Liouvillian~$\mathcal{L}$ in superoperator notation  (\eref{eq_liouville_lindblad}), in the block structure stated in \eref{eq_Lindbladmatrix_2} for the qubit and \textbf{a} one TLS and \textbf{b} two TLSs. We indicate non-zero real entries from $\Gamma_{\uparrow,\downarrow}^j$ in blue, non-zero imaginary entries from the couplings $g_k$ in red, and non-zero real and imaginary entries from various relaxation, dephasing, and frequency contributions in black.} \label{fig_liouvillian_matrices}
\end{center}
\end{figure}

First, we use this paragraph to define the order of the wave functions. The wave function $\ket{\psi}$ of the qubit and $n$ TLSs can be expressed as $\ket{\psi} = \sum_i c_i \ket{v_i}$, with the basis functions
\begin{align*}
    \begin{array}{rclrcl}
    		\ket{v_0} &=& \ket{1111\dots1},  &  \ket{v_1} &=& \ket{0111\dots1},\\
    		\ket{v_2} &=& \ket{1011\dots1}, &\ket{v_3} &=& \ket{0011\dots1},\\[-3pt]
            &\vdots& &&\vdots& \\
    	    \ket{v_{2^{n + 1} - 2}} &=& \ket{1000\dots0}, & \;\ket{v_{2^{n + 1} - 1}} &=& \ket{0000\dots0}.
    \end{array}
\end{align*}
Here, the first entry stands for the qubit, the second for the TLS $k = 1$, the third for the TLS $k = 2$, and so forth, with 0 and 1 denoting the ground and excited state, respectively.
In this basis, the density matrix reads
\begin{align*}
	\rho = \ket{\psi}\bra{\psi} = \begin{pmatrix}
		c_0c_0^* & c_0c_1^* & \cdots \\
		c_1c_0^* & c_1c_1^* &  \\
		\vdots & & \ddots
	\end{pmatrix}.
\end{align*}
To verify the structure and matrix elements in \eref{eq_Lindbladmatrix_2}
we state here the Liouvillian in superoperator notation following \cite{Schaller}. The trick is to write the density matrix as a tensor product $\rho \rightarrow \vec{\rho} = \sum \rho_{mn}\ket{m} \otimes \ket{n}$. It then follows $\mathbf{A}\rho \rightarrow (\mathbf{A} \otimes \mathds{1}) \vec{\rho}$ and similarly $\rho \mathbf{B} \rightarrow (\mathds{1} \otimes \mathbf{B}^T) \vec{\rho}$. We also need  $\mathbf{A}\rho \mathbf{B} \rightarrow (\mathbf{A} \otimes \mathbf{B}^T) \vec{\rho}$.
Since we have a symmetric Hamiltonian $H = H^T$ and real valued dissipators $L_\alpha = L_\alpha^*$ the Lindblad equation (\eref{eq_liouville_lindblad}) reads
\begin{align*}
	\vecev{\hspace*{1pt}\mathcal{L}\hspace*{3pt}} = &- \frac{i}{\hbar} (H \otimes \mathds{1} - \mathds{1} \otimes H) \\
	&+ \sum\limits_\alpha  L_\alpha \otimes L_\alpha - \frac{1}{2}(L_\alpha^T  L_\alpha \otimes \mathds{1} +  \mathds{1} \otimes L_\alpha^T L_\alpha).
\end{align*}
In this form, the matrix elements can easily be computed. For example, for the qubit coupled to two TLSs, one of the off-diagonal entries of $\mathbf{C}_\text{Z}$ originates from
\begin{align*}
	\vecev{\hspace*{1pt}\mathcal{L}\hspace*{3pt}}_{\!mn} &= \bra{100}\otimes\bra{010}\vecev{\hspace*{1pt}\mathcal{L}\hspace*{3pt}}\ket{101}\otimes\ket{011} =  \Gamma_\downarrow^{t_2}.
\end{align*}
Casting the tensor products in $\vecev{\hspace*{1pt}\mathcal{L}\hspace*{3pt}}$ and $\vec{\rho}$ in matrix and vector form gives the Liouvillian $\mathcal{L}$, which can then be structured as stated in \eref{eq_Lindbladmatrix_2} and depicted in \fref{fig_liouvillian_matrices}.

\section{Cross-relaxation between qubit and TLS in the presence of pure dephasing} \label{sec_stochastic_liouville}

In our derivation of the Solomon equations (\sref{sec_solomon}), we did not give a quantitative argument that allows for the use of the Markov approximation for the coherences, nor did we 
give any intuition for using the Lindblad equation in the first place. 
In the first part of this section, we therefore briefly review the classical derivation of the spin-lattice relaxation via the quantum stochastic Liouville equation \cite{Kubo1963Feb, Goldman2001Apr}. In the second part, we focus on the Markov approximation by presenting a detailed derivation for the system dynamics comprising the qubit and a single TLS, similar to \rcite{Solomon1955Jul}.

The discussion is greatly simplified when we assume that the qubit and the TLS are lossless. In this case, we can model the perturbation of the qubit and the TLS by a classical longitudinal noise source. The time-dependent Hamiltonian reads
\begin{align*}
H_\text{tot} &= \frac{\hbar\omega_\text{q}}{2}  \, \sigma^z_\text{q} + \frac{\hbar\omega_\text{t}}{2} \,  \sigma^z_\text{t} + \hbar g \, \sigma^x_\text{q} \sigma^x_\text{t} \,+ \\
&\qquad \quad + \hbar \eta_\text{q}(t)\sigma^z_\text{q}  + \hbar \eta_\text{t}(t) \sigma^z_\text{t}  ,
\end{align*}
where $\eta_\text{q,t}$ are classical noise sources with a certain power spectral density and zero mean. The Hamiltonian can be decomposed into two manifolds, namely the single-photon (flip-flop), and the two-photon (flip-flip) manifold. Since the derivation is analogous in both manifolds, we can focus on the single-photon manifold that gives the main contribution to the population transfer. Introducing the detuning $\delta = \omega_\text{t} - \omega_\text{q}$, one obtains
\begin{align*}
H &= \hbar \begin{pmatrix}
\frac{\delta}{2} & g\\ 
g & -\frac{\delta}{2}
\end{pmatrix} + \hbar \eta(t) \sigma_z
\end{align*}
with the mutual dephasing $\eta = \eta_\text{t} - \eta_\text{q}$. 
The evolution of the density matrix is given by the Liouville equation
\begin{align}
\dot{\rho} = - \frac{i}{\hbar} [H_0 + H_\eta(t), \rho] = (\mathcal{L}_0 + \mathcal{L}_\eta(t)) \rho = \mathcal{L}(t) \rho\label{eq_liouville}
\end{align}
with the solution $\rho(t) = \Phi(t) \rho(0)$. The propagator $\Phi(t)$ unfortunately cannot be written simply as a matrix exponential function, as $\mathcal{L}_0$ and $\mathcal{L}_\eta$ do not commute. However, as we are only interested in the averaged density matrix $\rho(t) = \langle \Phi(t)\rangle \rho(0)$, we wish to find an effective propagator $e^{\mathcal{L}_\mathrm{eff} t} \approx \langle \Phi(t)\rangle$ that is local in time and correctly captures the long-term dynamics of the density matrix. The general procedure of deriving the stochastic propagator $\langle \Phi(t)\rangle$ is discussed in great detail in \rcite{Kubo1963Feb}. We present thereof the relevant derivation and conditions that lead to a time-local master equation.

In short summary, the effective Liouvillian $\mathcal{L}_\mathrm{eff}$ gives a good approximation when the noise is sufficiently weak and incoherent such that the noise can be integrated independently on short timescales. For our system we can already surmise that for the correlation time $\tau_\text{c}$ of the noise it must hold that $\tau_\text{c} \ll 2 \pi / \Omega$, with $\Omega = \sqrt{4 g^2 + \delta^2}$ being the Rabi frequency. It follows that the coherence entries of the density matrix accumulate an additional random phase $\varphi$ within the correlation time. For isotropic noise, one then obtains on average the decoherence rate $\Gamma_2 =  (1 - \langle e^{i\varphi}\rangle) / \tau_\text{c}$.

More explicitly, we derive $\mathcal{L}_\text{eff}$ from second-order perturbation theory in the interaction picture \cite{Kubo1963Feb}. The density matrix in the interaction picture $\rho_\text{I}$ is defined by $\rho = e^{\mathcal{L}_0 t} \rho_\text{I}$ and governed by the equation of motion
\begin{align*}
\dot{\rho}_\text{I} = e^{-\mathcal{L}_0 t}\mathcal{L}_\eta(t)  e^{\mathcal{L}_0 t}\rho_\text{I} = \mathcal{L}_\text{I}(t) \rho_\text{I}.
\end{align*}
A formal solution $\rho_\text{I}(t) = U(t, t_0) \rho_\text{I}(t_0)$ is given by the Dyson series 
\begin{multline*}
U(t, t_0) = 1 + \int_{t_0}^t \text{d}t_1 \mathcal{L}_\text{I}(t_1) \\
+ \int_{t_0}^t \text{d}t_1 \mathcal{L}_\text{I}(t_1) \int_{t_0}^{t_1} \text{d}t_2 \mathcal{L}_\text{I}(t_2) + \dots \,,
\end{multline*}
so that the evolution of the averaged density matrix in the Schrödinger picture reads
\begin{align}
\dot{\rho} &= \mathcal{L}_0 \rho + e^{\mathcal{L}_0 t} \langle \dot{\rho}_\text{I}\rangle \nonumber \\
 &= \mathcal{L}_0 \rho + e^{\mathcal{L}_0 t} \frac{\text{d}}{\text{d}t} \langle U(t, t_0) \rangle \,\rho_\text{I} (t_0) \nonumber\\
           &= \mathcal{L}_0 \rho + e^{\mathcal{L}_0 t} \underbrace{\vphantom{(}\langle e^{-\mathcal{L}_0 t}\mathcal{L}_\eta(t)  e^{\mathcal{L}_0 t} \rangle}_{= 0} \rho_\text{I} (t_0) \nonumber \\[-5pt]
           &\quad + \! \int\limits_{t_0}^t \!\! \text{d}t' \langle \mathcal{L}_\eta(t) \underbrace{\vphantom{(}e^{-\mathcal{L}_0 (t - t')}}_{\approx 1} \mathcal{L}_\eta(t')  \underbrace{\vphantom{(} e^{-\mathcal{L}_0 (t-t')}}_{\approx 1} \rangle  \cdot e^{\mathcal{L}_0 t} \underbrace{\vphantom{(}\rho_\text{I} (t_0)}_{\approx \rho_\text{I} (t)}\nonumber \\
           &\quad + \, \dots \nonumber\\[0.1em]
           &\approx \left( \mathcal{L}_0 + \frac{1}{2} \int\limits_{-\infty}^\infty \langle \mathcal{L}_\eta(0) \mathcal{L}_\eta(\tau)\rangle\text{d}\tau \right) \rho = \mathcal{L}_\text{eff}\,\rho. \label{eq_deriv_stoch_liou}
\end{align}
In the above we made several assumptions.
First, we assumed that the noise is isotropic, and that it is sufficiently incoherent such that $\tau_\text{c} \ll 2 \pi / \Omega$ so that $t_0 \approx t - \tau_\text{c}$. Next, we supposed that the noise is weak enough so that we can approximate $\rho_\text{I}(t_0)$ by its current time $\rho_\text{I}(t)$ and that the Dyson series can be terminated at second order. Finally, we assume the noise to be stationary, and we extended the integral boundaries to infinity allowing us to link the decoherence rate $\Gamma_2 = 2 S_{\eta\eta}(0)$ with the power spectral density $S_{\eta\eta}(\omega)$ of the noise $\eta(t)$ \cite{Ithier2005Oct}.

Since the population $(\rho_{11} + \rho_{22}) \in [0, 1]$ in the single-photon manifold is conserved, we can rewrite \eref{eq_deriv_stoch_liou} as a special case of the three-dimensional Bloch equations \cite{Bloch1946Oct, Am-Shallem2015Nov}. Defining the polarization $z = \rho_{11} - \rho_{22}$, and directions $x = 2 \cdot \text{Re}(\rho_{21})$, and $y = 2 \cdot \text{Im}(\rho_{21})$, we obtain
\begin{align}
\frac{\text{d}}{\text{d}t}
\begin{pmatrix}
x \\ y \\ z
\end{pmatrix} &= - \begin{pmatrix}
\Gamma_2& \delta & 0 \\
- \delta& \Gamma_2 & 2g\\
0 & - 2g & 0
\end{pmatrix} \cdot
\begin{pmatrix}
x \\ y \\ z
\end{pmatrix}. \label{eq_lioville_dgl}
\end{align}
It immediately follows from the steady-state solution that, on average, both states of the single-photon manifold are equally populated. For a finite temperature, this steady-state solution contradicts the Boltzmann distribution when the qubit and the TLS have different energies. This result is typical for stochastic Liouville equations and a consequence of the real valued noise acting as an infinite temperature environment \cite{Goldman2001Apr}. The steady-state solution is a good approximation as long as the energy difference is small in comparison to the temperature.

What remains is to show under which conditions the coherent population transfer can be neglected. Solving the equation of motion requires finding the roots of the characteristic polynomial, which is cubic and cannot be easily factored. In the absence of a coherent population transfer, we expect the relaxation to be described by the real root with smallest absolute value, which will be shown in the following.

Fortunately, the structure of the characteristic polynomial of the Liouvillian in \eref{eq_lioville_dgl} allows for a good approximation in a wide parameter range.
We have
\begin{gather}
\lambda^3 - 2 \Gamma_2 \lambda^2 + (\Gamma_2^2 + 4g^2 + \delta^2) \lambda - 4 g^2 \Gamma_2 = 0,  \nonumber \\
\intertext{and by defining $\mu = \lambda / \Gamma_2$ we obtain}
\mu^3 - 2 \mu^2 + \frac{\Gamma_2^2 + 4g^2 + \delta^2}{\Gamma_2^2} \mu - 4 \frac{g^2}{\Gamma_2^2} = 0. \label{eq_cubic_root}
\end{gather}
Now, let $\mu_i$ denote the roots of the above polynomial and let $\mu_0$ denote the real root with lowest value in case that all roots are real. For the corresponding eigenvectors one finds $v_i = \begin{pmatrix} \delta \mu_i / (1 - \mu_i), & - \Gamma_2\mu_i,  & 2 g\end{pmatrix}^T $, which do not form an orthogonal basis as the Liouvillian in \eref{eq_lioville_dgl} is not a normal operator. 
Nevertheless, when $\mu_0 \rightarrow 0$ we can already surmise that this eigenvalue contributes dominantly to the relaxation. Unfortunately, it is not straightforward to give a quantitative answer, particularly because the eigenvalues can become degenerate. It is therefore very instructive to derive a few properties of the roots that allow for a better understanding of the resulting relaxation dynamics.

When $\Gamma_2^2 + \delta^2 \gg 4g^2 $ or, equivalently, when $\mu_0 \rightarrow 0$, the root $\mu_0$ can be approximated to first order,
\begin{align}
    \mu_0 \approx \frac{4g^2}{\Gamma_2^2 + 4g^2 + \delta^2}.  \label{eq_root_approx}
\end{align}Besides the negligible term $4g^2$ in the denominator and an overall factor of two, we obtain once again  the cross-relaxation rate (cf. \eref{eq_cross_relaxation}). The factor of 2 in the numerator simply results from the fact that $\lambda_0$ describes the qubit and TLS relaxation, whereas the cross-relaxation rate $\Gamma_\text{qt}$ is a transition rate between the qubit and TLS.

The next two roots $\mu_{1,2}$ can in principle be expressed with the help of $\mu_0$, but in doing so we do not gain any intuition and we would have to do a case analysis on the sign of the discriminant $D$ \footnote{Note that there are different sign conventions in the definition of the discriminant.} of the third-order polynomial. Instead, we follow a different approach. First, using the Routh-Hurwitz stability criterion, one can easily show that  $\mathrm{Re}(\mu_i) \in (0, 1]$.
Second, we have $\sum_i \mu_i = 2$. Thus, whenever $\mu_0 \rightarrow 0$ we have that $\mathrm{Re}(\mu_{1, 2}) \rightarrow 1$, and if the discriminant $D > 0$, the imaginary part is given by $\mathrm{Im}(\mu_{1, 2}) = \pm \sqrt{3/4 \mu_0^2 - \mu_0 + (4g^2 + \delta^2) / \Gamma_2^2}$.
As a side note, the value $\mu_i = 1$ is realized for one root if and only if $\delta = 0$. In this special case, the polynomial decomposes into $(\mu - 1)(\mu^2 - \mu + 4g^2/\Gamma_2^2) = 0$ with solutions
\begin{align*}
\mu_i = 1 \quad \lor \quad \mu_{j, k} = \frac{1}{2} \pm \sqrt{\frac{1}{4} - \frac{4g^2}{\Gamma_2^2}}.
\end{align*}

\begin{figure*}[t!]
\begin{center}
\includegraphics[scale=1.0]{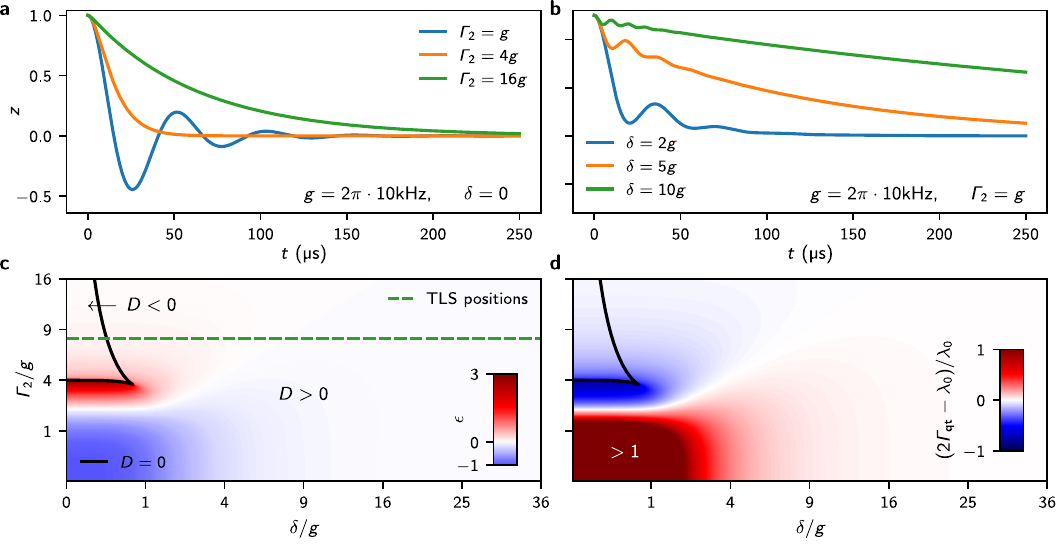}
\caption{\textbf{Qubit TLS cross-relaxation}. \textbf{a, b} Relaxation of the polarization $z$ in the qubit-TLS single-photon manifold for various decoherence rates $\Gamma_2$ (panel~a) and frequency detunings $\delta$ (panel~b). \textbf{c} Relative proportion $\epsilon$ (cf. \eref{eq_error}) of the initial polarization that does not relax via the cross-relaxation rate, as a function of the detuning and decoherence. Note the quadratic axes. The black curve shows where the discriminant $D$ is zero. For $D > 0$ the polarization can show oscillatory behavior. In the region where $\sqrt{\Gamma_2^2 + \delta^2} > 4g$, the incoherent cross-relaxation becomes the dominant relaxation mechanism. For reference, the green dashed line shows where the TLSs were located in the experiment assuming a mutual decoherence of $\Gamma_2 = \SI{0.5}{MHz}$. \textbf{d} Relative error in approximating the root $\lambda_0$ with the cross-relaxation rate.
} \label{fig:stochastic_liouville}
\end{center}
\end{figure*}

The relaxation dynamics following an initial condition $z_0$, $x_0$, and $y_0$, can be expressed as
\onecolumngrid
\begin{align}
    z(t) = \mathlarger{\sum}_{i=0}^2 \, (1 - \mu_i)\,\frac{(1 - \mu_{i+1})(1 - \mu_{i + 2}) 2g\Gamma_2\,x_0 + 2g\delta\,y_0 + \mu_{i+1}\mu_{i+2}\,\delta \Gamma_2\,z_0}{\delta \Gamma_2 (\mu_{i + 1} - \mu_{i})(\mu_{i + 2} - \mu_{i})} \cdot \mathlarger{e^{- \mu_i \Gamma_2 t}}, \label{eq_relax_analytic}
\end{align}
\twocolumngrid 
\noindent where we defined $\mu_{i + 3} = \mu_{i}$ cyclically. 
More compactly, we may write
\begin{align*}
     z(t) &= \sum_{i=0}^2 (u_i x_0 + v_i y_0 + w_i z_0) e^{-\mu_i \Gamma_2 t}
     = \sum_{i=0}^2 c_i e^{-\mu_i \Gamma_2 t}.
\end{align*}
Next, we need to show that $c_0 \rightarrow z_0$ for $\mu_0 \rightarrow 0$. In this case, after approximately $1 / \Gamma_2$, when the initial coherent onset has vanished, the remaining term is the slow decay of the polarization from $z_0$ at the rate $\mu_0 \Gamma_2$.
Using \eref{eq_cubic_root} and our knowledge of the roots we find
\begin{align*}
    u_0(\mu_0) &= \frac{\sqrt{\mu_0\mu_1\mu_2(1 - \mu_0)(1 - \mu_1)(1 - \mu_2)}}{(\mu_1 - \mu_0)(\mu_2 - \mu_0)} = \mathcal{O}(\sqrt{\mu_0}), \\
    v_0(\mu_0) &= \frac{\sqrt{\mu_0\mu_1\mu_2}}{(\mu_1 - \mu_0)(\mu_2 - \mu_0)} = \mathcal{O}(\sqrt{\mu_0}),\\
    w_0(\mu_0) &=\frac{(1 -\mu_0)\mu_1\mu_2}{(\mu_1 - \mu_0)(\mu_2 - \mu_0)} = 1 - \mathcal{O}(\mu_0).
\end{align*}
We see here that the incoherent relaxation with $e^{- \mu_0 \Gamma_2 t}$ with the prefactor $c_0$ is well behaved when $\mu_0 \rightarrow 0$.

In order to also be concrete for the coherent population transfer with $e^{- \mu_{1, 2} \Gamma_2 t}$, we must handle the prefactors $c_{1, 2}$, which may diverge when the eigenvalues become degenerate. 
One way to circumvent the divergence is to calculate the area $A$ under the relaxation curve by integrating \eref{eq_relax_analytic}, which can then be compared to the corresponding exponential decay with the rate $\mu_0 \Gamma_2$. In doing so, we get
\begin{align}
 A(x_0, y_0, z_0) &= \int\limits_0^\infty \!z(t) \text{d}t \, =\mathlarger{\int}\limits_0^\infty \sum_{i = 0}^2 c_i e^{-\mu_i \Gamma_2 t} \text{d}t \nonumber \\[3pt]
  &= \frac{\delta x_0 +\Gamma_2 y_0}{2g \Gamma_2} + \frac{\delta^2 + \Gamma_2^2}{4g^2 \Gamma_2} \, z_0. \label{eq_A}
\end{align}
Interestingly, the inverse of the prefactor of $z_0$ is exactly the twofold cross-relaxation rate $\Gamma_\text{qt}$ that was derived in the main text (\eref{eq_cross_relaxation}). We therefore conclude that $\Gamma_\text{qt}$ is the average rate when the system starts in an incoherent state, meaning that $x_0 = y_0 = 0$. 
The relative proportion of $z_0$ that decays coherently is
\begin{align}
    \epsilon &= \left(A(0, 0, z_0)  - \int\limits_0^\infty \! z_0e^{- \mu_0 \Gamma_2 t} \,\text{d}t \right) \Big/  \int\limits_0^\infty \! z_0e^{- \mu_0 \Gamma_2 t}\, \text{d}t \nonumber \\
    &= \mu_0 \left(\frac{\mu_1 + \mu_2}{\mu_1\mu_2} - 1\right), \label{eq_error}
\end{align}
which approaches zero for $\mu_0 \rightarrow 0$.
Similarly, one might wonder to what extend an initial coherence is converted into polarization.
With $z_\text{max} = \rho_{11} + \rho_{22}$ we find the maximal area to be
\begin{align*}
	\max_{x_0, y_0} A(x_0, y_0, z_0) = \frac{\sqrt{z_\text{max}^2 - z_0^2}\sqrt{\Gamma_2^2 + \delta^2}}{2 g \Gamma_2},
\end{align*}
which should be compared to $\mu_0 \Gamma_2$ to give the effectively reached polarization
\begin{align*}
	z_\text{eff} = \mu_0 \frac{\sqrt{z_\text{max}^2 - z_0^2}\sqrt{\Gamma_2^2 + \delta^2}}{2 g},
\end{align*}
approaching zero for $\mu_0 \rightarrow 0$. 
In summary, we conclude that if  $\sqrt{\Gamma_2^2 + \delta^2} \gg 4g$, the relaxation of the population is governed by
\begin{align}
    \dot{z} = -\mu_0 \Gamma_2 z. \label{eq_z_final}
\end{align}
In \fref{fig:stochastic_liouville} we show various solutions of the Liouville equation using \eref{eq_relax_analytic}, the relative proportion $\epsilon$, and the relative approximation error of the root $\lambda_0$.

For sake of completeness, we can conveniently derive again the Solomon equations for our current simple scenario. From \eref{eq_z_final}, it follows for the qubit
\begin{align*}
	\dot{\rho}_{22} = - \dot{\rho}_{11} = - \mu_0 \Gamma_2 \frac{\rho_{22} - \rho_{11}}{2}.
\end{align*}
Inserting the definition of the probabilities $p_\text{q} = \rho_{00} + \rho_{22}$ and $p_\text{t} = \rho_{00} + \rho_{11}$ and for simplicity neglecting contributions from two-photon processes,
we obtain the Solomon equations 
\begin{align*}
     \quad \dot{p}_\text{q} &= - \frac{\mu_0 \Gamma_2}{2} p_\text{q} + \frac{\mu_0 \Gamma_2}{2} p_\text{t}, \\
    \dot{p}_\text{t} &= - \frac{\mu_0 \Gamma_2}{2} p_\text{t} + \frac{\mu_0 \Gamma_2}{2} p_\text{q}, 
\end{align*}
with the cross-relaxation rate
\begin{align}
    \Gamma_\text{qt} = \frac{\mu_0 \Gamma_2}{2} \approx \frac{2 g^2 \Gamma_2}{\Gamma_2^2 + \delta^2}.
\end{align}

\section{Quantum jumps from stochastic Schrödinger equation} \label{sec_SSE}

As we have shown in the main text, the measured quantum jump statistics can also be explained by an unraveling of the Lindblad master equation, which includes the measurement backaction. 
The unraveling of master equations is not unique, but can sometimes be motivated by the underlying physical processes \cite{Wiseman1993Mar, Minev2019Jun}. 
A successful modeling of the measured quantum jump statistics may therefore distinguish different unravelings and could reveal a deeper understanding of the qubit environmental interaction.

Following the discussion in the main text (\sref{sec_quantum_jumps}), we will first consider a  closed system consisting of the qubit and the discrete TLS environment under the presence of pure dephasing. As in \aref{sec_stochastic_liouville}, we may assume a diffusive process for the dephasing, meaning that the qubit and the TLSs accumulate a random phase during their coarse-grained evolution. The state diffusion may therefore be modeled by the stochastic Schrödinger equation.
To speed up the simulation, we perform the rotating-wave approximation for the $\sigma_x^\text{q} \sigma_x^\text{k}$ qubit TLS interaction in \eref{eq_hamiltonian}, so that the integration time step $\text{d} t = 1/32\,\si{\micro s}$ can be kept on an acceptable level in the interaction picture. Due to the longitudinal noise, transformation to the interaction picture can be performed exactly via
\begin{align*}
    \ket{\psi_\text{I}(t)}
    &= \prod\limits_j e^{i \left(\frac{\omega_j t}{2} + \sqrt{\frac{\Gamma^j_\varphi}{2}} W_t^j\right) \sigma_z^j} \ket{\psi(t)},
\end{align*}
with $W^j_t$ denoting independent Wiener processes. 
In addition, the longitudinal noise allows separating the simulation into the excitation manifolds.
Integration of the wave function is performed by
\begin{align*}
    \ket{\psi_\text{I}(t + \text{d}t)} = \ket{\psi_\text{I}(t)} - \frac{i}{\hbar} H_\text{I}(t) \ket{\psi_\text{I}(t)} \text{d}t,
\end{align*}
after which we normalize the wave function. The instantaneous projective qubit measurements are performed at $t_\text{rep} = \SI{2}{\micro s}$ intervals. For the qubit we use the measured dephasing rate $\Gamma_\varphi^\text{q} = \SI{0.5}{MHz}$. For the unknown dephasing of the TLSs, we use $\Gamma_\varphi^\text{t} = (\text{0.1 or 1.0})\,\si{MHz}$.

\begin{figure*}[ht!]
\begin{center}
\includegraphics[scale=1.0]{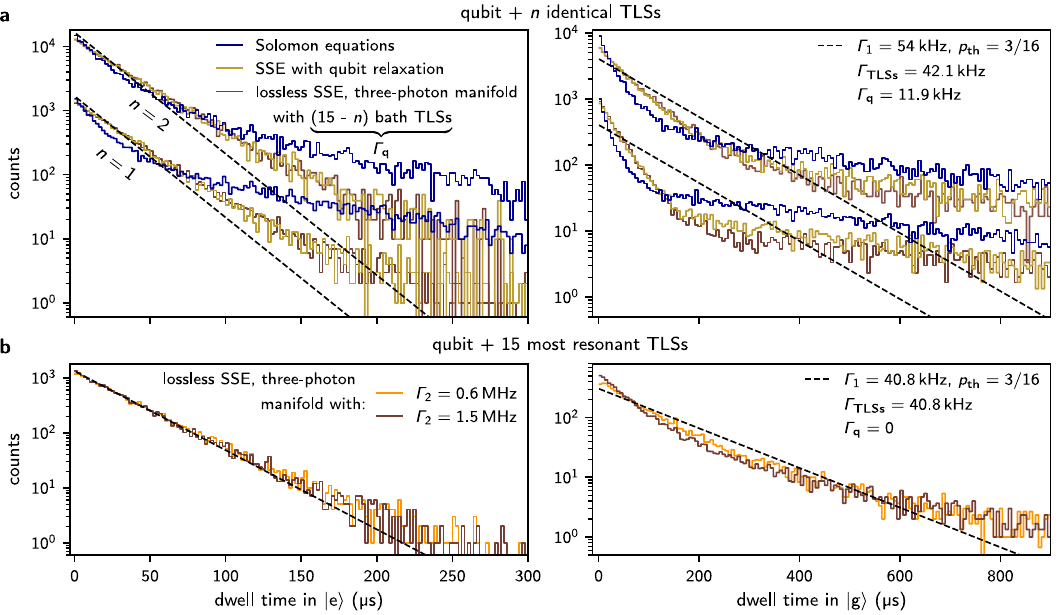}
\vspace{-4mm}
\caption{\textbf{Comparison of various simulated quantum jump statistics.} 
\textbf{a} Simulated quantum jump statistics of the qubit dwell times in $|\text{e}\rangle$ (left panels) and $|\text{g}\rangle$ (right panels), respectively, assuming that the qubit is coupled strongly to $n = \text{1 or 2}$ TLSs, such that their cross-relaxation adds up to the total measured cross-relaxation $\Gamma_\text{TLSs}$. The distributions for $n = 2$ were shifted upwards by a factor 10 for better visibility.
In the lossless SSE simulation, the qubit couples in addition 
weakly to (15 - n) TLSs which model the intrinsic qubit relaxation $\Gamma_\text{q}$. In all simulations, therefore, the thermal population was therefore set to $p_\text{th} = 3 / 16$.  For similar measured distributions, see \rcite{Spiecker2024}. \textbf{b} Lossless SSE simulation for the qubit and the 15 most resonant TLSs using the measured experimental parameters (\frefadd{fig_loglog}{b}), while assuming the system state trapped in the three-photon manifold. We simulate with the dephasing rates $\Gamma_\varphi^\text{q} = \SI{0.5}{MHz}$ and $\Gamma_\varphi^\text{t} = (\text{0.1 and 1.0})\,\si{MHz}$. For the higher coherence, a non-Markovian onset can be seen in the ground-state statistics. The histograms are generated as described in \fref{fig_intro}, except that all distributions are based on 20000 counts.} \label{fig_quantum_jumps_SSE}
\vspace{-4mm}
\end{center}
\end{figure*}

As discussed and shown in \fref{fig_backaction} in the main text, under the presence of stroboscopic qubit measurements the system state will eventually be trapped in one of the photon manifolds when the system is closed. 
One might wonder if in this case a non-exponential quantum jump behavior can still be observed. 
We decided to investigate a system consisting of 16 elements trapped in the three-photon manifold, which corresponds to a thermal population of $p_\text{th} = 3 / 16$, similar to the experiment.
To illustrate that the SSE and the Solomon equations lead to distinct non-exponential distributions and to validate the SSE with qubit relaxation, we perform simulations where the entire measured cross-relaxation $\Gamma_\text{TLSs}$ originates from only one or two long-lived TLSs in resonance with the qubit and with $\Gamma_\varphi^{\text{t}_k} = \SI{1.0}{MHz}$. The remaining 14 or 13 weakly coupled TLSs are used to emulate the measured intrinsic qubit loss. We assume for them $\Delta = 0$, and $\Gamma_\varphi^{\text{t}_k} = \SI{1.0}{MHz}$ as well. 
The simulation in the three-photon manifold provides enough excitations to populate the qubit and the strongly coupled TLSs. The qubit will therefore experience a fluctuating temperature (s. \eref{eq_equilibrium_pop_TLSs}) resulting from the random walk on the hypersphere. 
The various simulation results are depicted in \frefadd{fig_quantum_jumps_SSE}{a}. Interestingly, simulations with the lossy SSE, which incorporates the qubit relaxation, yield the same statistics, whereas the classical behavior expected from the Solomon equations differs clearly. The highly non-exponential distributions predicted by the SSE are also observed experimentally for one TLS being strongly coupled to the qubit~\cite{Spiecker2024}.

In addition, we depict in \frefadd{fig_quantum_jumps_SSE}{b} the quantum jump statistics of the experimental situation for the qubit and the 15 most resonant TLSs. Qualitatively, the experimentally observed non-exponential distribution can be reproduced. The non-Markovian onset for $\Gamma_2 = \SI{0.6}{MHz}$ visible in the ground-state statistics, which is absent in the experiment, disappears when the decoherence time becomes smaller than the measurement repetition time. 
For this reason, we adhered to $\Gamma_\varphi^{\text{t}_k} = \SI{1.0}{MHz}$ in all other simulations. A comparison with the Solomon equations is precluded, due to the absence of qubit relaxation.

Instead of including additional qubit environments in the simulation, transitions between the excitation manifolds may also be induced by modeling the qubit relaxation with jump operators. They can be interpreted as measurements on the qubit's intrinsic environments. In the spirit of the previous simulations, we decided to use again a diffusive SSE. In this case, integration of the phase noise and of the interaction Hamiltonian can be performed as before. For the diffusive relaxation, we have in addition \cite{Gisin1992Nov, Castin1993Jun, breuer2002theory}
\begin{align*}
    &\ket{\psi_\text{I}(t + \text{d}t)} = \ket{\psi_\text{I}(t)} - \frac{i}{\hbar} H_\text{int}(t) \ket{\psi_\text{I}(t)} \text{d}t \\
    & \quad + \sum\limits_\alpha \left(\langle L_\alpha^\dagger\rangle L_\alpha - \frac{1}{2} L_\alpha^\dagger L_\alpha  -  \frac{1}{2} \langle L_\alpha^\dagger\rangle \langle L_\alpha \rangle \right) \ket{\psi_\text{I}(t)} \text{d}t \\
    &\quad + \sum\limits_\alpha \left(L_\alpha - \langle L_\alpha \rangle \right) \ket{\psi_\text{I}(t)}\frac{\text{d}W_\alpha}{\sqrt{2}},
\end{align*}
with $\langle L_\alpha \rangle = \langle\psi_\text{I}| L_\alpha | \psi_\text{I}\rangle$, the qubit jump operators $L_1 = \sqrt{\Gamma_\downarrow^\text{q}} \sigma_\text{q}^-$ and $L_2 = \sqrt{\Gamma_\uparrow^\text{q}} \sigma_\text{q}^+$, and $\text{d}W_\alpha$ independent complex Wiener increments. Each step is followed by a normalization of the wave function.

\section{Approximating the inverse of the coherences matrix} \label{sec_matrix_inverse}

For the derivation of the Solomon equations, we need to compute $\mathbf{C}^{-1}$, where we can make use of the block-wise inversion formula. 
We write $\mathbf{C}$ as
\begin{align*}
    \mathbf{C} = \begin{pmatrix} \mathbf{C}_\text{ZD} &  \mathbf{S}_\text{ZD}^T \\ 
    \mathbf{S}_\text{ZD} &  \mathbf{C}_\text{R} \end{pmatrix},
\end{align*}
with index ZD denoting the joint matrices. 
Under the assumption that the inverse matrices $\mathbf{C}_\text{ZD}^{-1}$ and $(\mathbf{C}_\text{R} - \mathbf{S}_\text{ZD}\mathbf{C}_\text{ZD}^{-1} \mathbf{S}_\text{ZD}^T)^{-1}$ exist, the upper left block of the inverse matrix \raisebox{0pt}[9pt][0pt]{$\mathbf{C}^{-1}$} is
\begin{gather*}
    \mathbf{C}_\text{ZD}^{-1} +  \mathbf{C}_\text{ZD}^{-1} \mathbf{S}_\text{ZD}^T\left(\mathbf{C}_\text{R} - \mathbf{S}_\text{ZD}\mathbf{C}_\text{ZD}^{-1} \mathbf{S}_\text{ZD}^T\right)^{-1} \mathbf{S}_\text{ZD}\mathbf{C}_\text{ZD}^{-1}.
\end{gather*}
Next, we can use the Neumann series to obtain a Taylor expansion for the inverse of a matrix. For a given matrix $\mathbf{A}$, with the decomposition $\mathbf{A} = \mathbf{D} - \mathbf{O}$ with $\mathbf{D}$ being diagonal, it holds that
\begin{align*}
    \mathbf{A}^{-1} = \mathbf{D}^{-1} + \mathbf{D}^{-1} \mathbf{O} \mathbf{D}^{-1} + \mathbf{D}^{-1} \mathbf{O} \mathbf{D}^{-1} \mathbf{O} \mathbf{D}^{-1} + \dots,
\end{align*}
as long as $\Vert \mathbf{D}^{-1/2} \mathbf{O}  \mathbf{D}^{-1/2} \Vert < 1$. With this expansion, we find in lowest order in the diagonal entries
\begin{multline*}
    \mathbf{C}^{-1} = \mathbf{D}_\text{ZD}^{-1} +  \mathbf{D}_\text{ZD}^{-1} \mathbf{O}_\text{ZD} \mathbf{D}_\text{ZD}^{-1} + \\
    + \mathbf{D}_\text{ZD}^{-1} \mathbf{S}_\text{ZD}^T \mathbf{D}_\text{R}^{-1} \mathbf{S}_\text{ZD}\mathbf{D}_\text{ZD}^{-1} + \mathcal{O}\left(\mathbf{D}^{-4}\right).
\end{multline*}

\section{Proof of the transformation in \eref{eq_coord_transf_rate_eq}}\label{sec_proof_solomon}

For more TLSs it is expedient to express the rate equation $\mathcal{L}_\text{D}$ with the help of spin creation and annihilation operators. These operators obey the anticommutation relations 
\begin{gather*}
	\{\sigma_j^-, \sigma_j^+\} = 1, \quad  \{\sigma_j^-, \sigma_j^-\} = \{\sigma_j^+, \sigma_j^+\} = 0 
\end{gather*}
for the same element, while for different elements  $j \neq i$ they commute with each other:
\begin{gather*}
	[\sigma_j^-, \sigma_i^-] = [\sigma_j^+, \sigma_i^+] = [\sigma_j^-, \sigma_i^+] = 0.
\end{gather*}
Using these operators, the non-Hermitian rate equation $\mathcal{L}_\text{D}$ can now be expressed for an arbitrary number of TLSs. For brevity, we will focus on the scenario in which the TLSs are not interacting with each other. We have
\begin{align}
\begin{aligned}
	\mathcal{L}_\text{D} &= \sum\limits_{j = 0}^n \Gamma^j_\downarrow (\sigma_j^- - \sigma_j^+ \sigma_j^-) + \sum\limits_{j = 0}^n \Gamma^j_\uparrow (\sigma_j^+ - \sigma_j^- \sigma_j^+) \\[-3pt]
	& \qquad  + \sum\limits_{j = 1}^n \Gamma_\text{qt}^{\delta_j} (\sigma_j^+ \sigma_0^- - \sigma_j^- \sigma_j^+ \sigma_0^+ \sigma_0^-  \\[- 8pt] 
	& \qquad \qquad \qquad + \sigma_j^- \sigma_0^+ - \sigma_j^+ \sigma_j^- \sigma_0^- \sigma_0^+) \\
	& \qquad + \sum\limits_{j = 1}^n \Gamma_\text{qt}^{\sigma_j} (\sigma_j^+ \sigma_0^+ - \sigma_j^+ \sigma_j^- \sigma_0^+ \sigma_0^- \\[- 8pt]
	& \qquad \qquad \qquad + \sigma_j^- \sigma_0^- - \sigma_j^- \sigma_j^+ \sigma_0^- \sigma_0^+), \label{eq_rate_creation_annihilation}
\end{aligned}
\end{align}
where we use the index $j = 0$ for the qubit and $j > 0$ to denote the TLSs.
The covariant vectors $\bra{v_i}$ such that $p_i = \bra{v_i} \rho_\text{D}$ as well as $1 = \bra{1} \rho_\text{D}$ can be created via
\begin{align*}
	\bra{v_i} &= \bra{0} \sigma_i^- \prod\limits_{j \neq i} (1 + \sigma_j^-), \\
	\bra{1} &= \bra{0}\prod\limits_j (1 + \sigma_j^-).
\end{align*}
The next step is to show that these vectors are mapped onto each other when they are applied to the rate equation in \eref{eq_rate_creation_annihilation} from the left.
The first sum simply results in $\bra{v_i}\mathcal{L}_\text{D} = - \Gamma_\downarrow^i \bra{v_i} + \cdots$. In a similar way,  the second sum vanishes for $j \neq i$. However, for $j = i$ the simplification is more intricate. Here, we can show
\begin{align*}
	&\bra{0} \sigma_i^-(\sigma_i^+ - \sigma_i^- \sigma_i^+) \cdots = \bra{0} \sigma_i^-(\sigma_i^+ - 1 + \sigma_i^+ \sigma_i^-) \cdots \\
	&\quad= - \bra{v_i} + \bra{0} (1 + \sigma_i^- - 1)(\sigma_i^+ + \sigma_i^+\sigma_i^-)\cdots \\
	&\quad= - \bra{v_i} + \bra{0}(1 + \sigma_i^-)(\sigma_i^+ + 1 - \sigma_i^-\sigma_i^+)\cdots - \cancel{\bra{0} (\sigma_i^ + \cdots} \\
	&\quad= - \bra{v_i} + \bra{1} - \underbrace{\bra{0}(1 + \sigma_i^-)(\sigma_i^+ - \sigma_i^-\sigma_i^+)\cdots}_{\cancel{\bra{0}\sigma_i^+\cdots}}\\[-13pt]
	&\quad= - \bra{v_i} + \bra{1},
\end{align*}
thus, $\bra{v_i}\mathcal{L}_\text{D} = \dots + \Gamma_\downarrow^i (\bra{1} - \bra{v_i}) + \cdots$. The third sum yields \vspace*{-1em}
\begin{align*}
	\bra{v_0}\mathcal{L}_\text{D} &= \cdots + \sum\limits_{j = 1}^{n} \Gamma_\text{qt}^{\delta_j} \left(\bra{v_j} - \bra{v_0}\right) + \cdots, \\
	\bra{v_i}\mathcal{L}_\text{D} &= \cdots + \Gamma_\text{qt}^{\delta_i} \left(\bra{v_0} - \bra{v_i}\right) + \cdots \quad \text{for $i > 0$}.
\end{align*}
Finally, the fourth sum goes to \vspace*{-0.2em}
\begin{align*}
	\bra{v_0}\mathcal{L}_\text{D} &= \cdots + \sum\limits_{j = 1}^{n} \Gamma_\text{qt}^{\sigma_j} \left(\bra{v_j} - \bra{v_0} + \bra{1}\right),  \\
	\bra{v_i}\mathcal{L}_\text{D} &= \cdots + \Gamma_\text{qt}^{\sigma_i} \left(\bra{v_0} - \bra{v_i} + \bra{1}\right)\quad \text{for $i > 0$}.
\end{align*}
The last two equations can be incorporated into the previous equations by redefining the rates as stated in the main text. Thus, we have $\bra{v_i} \mathcal{L}_\text{D} = \sum_j \bar{\mathbf{A}}_{ij} \bra{v_j} + \bar{\mathbf{\Gamma}}_{\uparrow, i} \bra{1}$.
Consequently, for any evolution following $\dot{\rho}_\text{D} = \mathcal{L}_\text{D}\rho_\text{D}$ we obtain the Solomon equations for the expectation values of the populations:
\begin{align*}
	\dot{p}_i = \bra{v_i}\dot{\rho}_\text{D} = \bra{v_i}\mathcal{L}_\text{D}\rho_\text{D} = \sum_j \bar{\mathbf{A}}_{ij} p_j + \bar{\mathbf{\Gamma}}_{\uparrow, i}.
\end{align*}

\section{Analytic solutions and approximations} \label{sec_solomon_solutions}

\begin{center}
	\bf \small 1. The case of identical cross-relaxation rates
\end{center}
As discussed in the main text, the system dynamics must be governed by the two eigenvectors $\mathbf{v}_{0,2} = \begin{pmatrix} x_{0,2} & 1 \dots 1 \end{pmatrix}^T$ and corresponding eigenvalues $\lambda_{0,2} = - \Gamma_\text{qt} x_{0,2}   + \Gamma_\text{t} + \Gamma_\text{qt}$ with 
\begin{align*}
- 2 \Gamma_\text{qt} x_{0,2} = \underbrace{\vphantom{\sqrt{(~)^2 + 4 n\Gamma_\text{qt}^2}} \Gamma_\text{q} + (n - 1) \Gamma_\text{qt} - \Gamma_\text{t}}_{(~)} \pm \underbrace{\sqrt{(~)^2 + 4 n\Gamma_\text{qt}^2}}_{\sqrt{\phantom{\Gamma}}}.
\end{align*}
Given the initial out-of-equilibrium populations of the qubit $p_{\text{q},0}^* := p_\text{q}^*(t=0)$ and of the TLSs $p_{\text{t},0}^* := p_\text{t}^*(t=0)$, it holds
\begin{align*}
\mathbf{p}^*_0 = \frac{p_{\text{t},0}^* - q}{2} \mathbf{v}_0 + \frac{p_{\text{t},0}^* + q}{2} \mathbf{v}_2 
\end{align*}
with $q = (2 \Gamma_\text{qt} p_{\text{q},0}^* + (~)\,p_{\text{t},0}^*) / \sqrt{\phantom{\Gamma}}$,
and the qubit and TLS dynamics finally read
\begin{align}
    \begin{array}{r@{\,}c@{}l@{}c@{}l}
    	p_\text{q}^*(t) =& \displaystyle{\frac{x_0}{2}} &(p_{\text{t},0}^* - q) e^{- \lambda_0 t} \,+\,&\displaystyle{\frac{x_2}{2}}& (p_{\text{t},0}^* + q) e^{- \lambda_2 t}, \\[2ex]
    	p_\text{t}^*(t) =& \displaystyle{\frac{1}{2}} &(p_{\text{t},0}^* - q) e^{- \lambda_0 t} \,+\, &\displaystyle{\frac{1}{2}}& (p_{\text{t},0}^* + q) e^{- \lambda_2 t}. 
    \end{array} \label{eq_relaxation_identical_rates}
\end{align}
The time-dependent transition rates and 
the equilibrium population can now be easily computed using \eqsref{eq_up_rate}, \ref{eq_down_rate} and \ref{eq_equilibrium_pop}.\\
 
\begin{center}
	\bf \small 2. The case of distributed cross-relaxation rates
\end{center}
The Pick function of the distribution \eref{eq_rates_quadratic} may first be brought into the form
\begin{align}
f(\lambda) &= \Gamma_\text{q} - \Gamma_\text{t} - \frac{a}{a / \lambda'}  - \sum\limits_{k = 1}^\infty \frac{a}{a / \lambda'- k^2} \label{eq_pick_pole}
\end{align}
with $\lambda' = \lambda - \Gamma_\text{t}$.
Here, we see the nature of the Pick function: it is a meromorphic function of $z$ in the complex plane, with $z = a / \lambda' = a / (\lambda - \Gamma_\text{t})$. 
The sum in \eref{eq_pick_pole} can be expressed in closed form. It holds
\begin{align}
\frac{f(z)}{a} &= \gamma - \frac{1}{2z} - \frac{\pi}{2 \sqrt{z}} \cot \pi \sqrt{z},
\end{align}
where we introduced $\gamma = (\Gamma_\text{q} - \Gamma_\text{t}) / a$.
The normalization can now conveniently be calculated via
\begin{align}
    \Vert v_m \Vert^2 &= - f'(\lambda)\Big|_{\lambda = \lambda_m} = z^2 \left.\frac{\partial}{\partial z} \frac{f(z)}{a} \right|_{z = z_m} \nonumber \\
    &= \frac{1}{2} + \left(\frac{\pi^2}{4} - \frac{\gamma}{2}\right) z_m + \gamma^2 z_m^2. \label{eq_pick_zm}
\end{align}
Here, we see that $\Vert v_m\Vert^2$ is a continuous function of $z_m$, which allows the use of simple approximations for the roots $z_m$ in the next step. 

So far our analysis is still exact. We are left with the evaluation of the sum in \eref{eq_pq_analytical}, which becomes an integral on long timescales. This motivates us to interpret $z$ as a continuous function of $m$.
The structure of \eref{eq_chauchy} with the rates $\Gamma_\text{qt}^k$, given by \eref{eq_rates_quadratic}, suggests
\begin{align*}
\lambda_m &= \frac{a}{(m + \frac{1}{2} + \delta(m))^2} + \Gamma_\text{t} \\
\Rightarrow \quad z(m) &= \left(m + \frac{1}{2} + \delta(m)\right)^2, 
\end{align*}
where we defined the deviation $\delta(m) \in (-1/2, 1/2)$.

As a side note, a very good approximation can be found for the $m^\text{th}$ root when the Pick function is approximated piecewise by its surrounding poles. We will make use of this later for the analysis of a more general rate distribution. Here, we obtain
\begin{align}
    \begin{aligned}
    z(m) &\approx m^2 + m + 1/\gamma + 1/2 \\
    					 &\qquad - \text{sgn}(\gamma) \sqrt{m^2 + m + 1/\gamma^2 + 1/4},
    \end{aligned}  \label{eq_approx_roots}
\end{align}
which becomes $z(m) = m^2 + m + 1 / 2$ for $\gamma = 0$. 
At first glance, when the derivative $f'$ is also approximated by the surrounding poles, insertion of \eref{eq_approx_roots} gives the correct limit $\Vert v_m \Vert^2 \rightarrow \gamma^2 m^4$ for large $m$ and $\gamma^2 > 0$. However, when checking $\gamma^2 = 0$, one obtains the slightly inaccurate limit $\Vert v_m \Vert^2 \rightarrow 2 m^2$ instead of $\Vert v_m \Vert^2 \rightarrow \beta m^2 = (\pi^2 / 4) m^2$ according to \eref{eq_pick_zm}.

With the function $z(m)$ we have everything needed to evaluate the sum in \eref{eq_pq_analytical} on long timescales. We begin with the simpler scenario in which $\gamma^2 = 0$, which essentially describes the qubit relaxing into the TLS environment. Defining $y(m) = \left(x(m) + \frac{1/2 + \delta(m)}{\sqrt{at}}\right)$ and $x(m) = \frac{m}{\sqrt{at}}$, we have
\begin{align*}
&\sum\limits_{m = 0}^\infty \frac{e^{- \lambda'_m t}}{\Vert v_m \Vert^2} = \frac{1}{\sqrt{at}}\sum\limits_{m = 0}^\infty \frac{e^{- \frac{1}{y(m)^2}}}{\frac{1/2}{at} + \beta{\kern0.1em}y{\kern0.02em}(m)} \frac{1}{\sqrt{at}} \\
&\;\; \approx \frac{1}{\sqrt{at}} \int\limits_{0}^\infty  \frac{e^{- \frac{1}{y\left(m(x)\right)^2}}}{\frac{1/2}{at} + \beta{\kern0.125em}y{\kern-0.125em}\left(m(x)\right)} \text{d}x  
\approx \frac{1}{\sqrt{at}} \int\limits_0^\infty \frac{e^{- \frac{1}{x^2}}}{\frac{1/2}{at} + \beta x^2} \text{d}x \\
&\;\; \approx \frac{1}{\sqrt{at}} \int\limits_0^\infty \frac{e^{- \frac{1}{x^2}}}{\beta x^2} \text{d}x = \frac{\sqrt{\pi}}{2 \beta}\frac{1}{(at)^{1/2}},
\end{align*}
where at first the sum was approximated by an integral, requiring $\sqrt{at} \gg 1$, which also allows the second approximation $(1 / 2 + \delta) / \sqrt{at} \approx 0$, since $|\delta| < 1/2$. In the last step, the denominator was approximated, which is valid when $\sqrt{at} \gg 1 / \sqrt{2\beta}$.

The scenario $\gamma^2 > 0$ can be treated similarly. We have
\begin{align*}
&\sum\limits_{m = 0}^\infty \frac{e^{- \lambda'_m t}}{\Vert v_m \Vert^2} \approx \frac{1}{(at)^{3/2}}\int\limits_0^\infty\frac{e^{- \frac{1}{x^2}}}{\frac{1/2}{(at)^2} + \frac{\beta}{at} x^2 + \gamma^2 x^4} \frac{1}{\sqrt{at}} \\
&\qquad\approx \frac{1}{(at)^{3/2}} \int\limits_0^\infty \frac{e^{- \frac{1}{x^2}}}{\gamma^2 x^4} \text{d}x = \frac{\sqrt{\pi}}{4\gamma^2}\frac{1}{(at)^{3/2}},
\end{align*}
which in addition to the previous scenario requires $\sqrt{at} \gg \sqrt{\beta / \gamma^2}$. The final limit relaxation behaviors are stated in the main text.

For the generalized distribution in \eref{eq_gamma_d} we find
analogously 
\begin{align*}
f(\lambda) &= \Gamma_\text{q} - \Gamma_\text{t} - \frac{a}{a / \lambda'}  - \sum\limits_{k = 1}^\infty \frac{a}{a / \lambda'- k^d}.
\end{align*}
Unfortunately, to our knowledge, this function cannot be expressed in a closed form, nor can its derivative be expressed via the function itself. For integer values of $d$, one can express $f$ as a sum of digamma functions, which for even integers can be rewritten as a sum of cotangents. In \aref{sec_special_solutions}, we show the relevant cases $d = 3$ and $d = 4$. We will therefore proceed as discussed before and approximate $f$ piecewise by its surrounding poles.
For $\gamma^2 > 0$ we then find $\Vert v_m \Vert^2 \rightarrow \gamma^2 (m + 1/2 + \delta(m))^{2 d}$ for large $m$ in congruence with \eref{eq_pick_zm}. For $\gamma^2 = 0$ one finds in leading order $\Vert v_m \Vert^2 \rightarrow 8 / d^2 m^2$, which surprisingly is always quadratic in $m$ and not with the power of $d$, as one might surmise from \eref{eq_pick_zm}. Note, as discussed before, the prefactor $8 / d^2$ is only an approximation. For instance, for $d = 3$ the correct prefactor is  $4\pi^2/27$ and for $d = 4$ one finds $\pi^2/8$ (cf. \aref{sec_special_solutions}). We therefore continue by reintroducing the prefactor $\beta$ via $\Vert v_m \Vert^2 \rightarrow \beta m^2$ for large $m$.  
As was done before, we have for $\gamma^2 = 0$
\begin{align*}
&\sum\limits_{m = 0}^\infty \frac{e^{- \lambda'_m t}}{\Vert v_m \Vert^2} \approx \frac{1}{(at)^{1 / d}}\sum\limits_{m = 0}^\infty \frac{e^{- \frac{1}{y(m)^d}}}{\beta \frac{m^2}{(at)^{2 / d}}} \frac{1}{(at)^{1 / d}} \\
&\;\; \approx \frac{1}{(at)^{1 / d}} \int\limits_{0}^\infty  \frac{e^{- \frac{1}{x^d}}}{\beta x^2} \text{d}x
= \frac{\Gamma(1 + \frac{1}{d})}{\beta} \frac{1}{(at)^{1 / d}}
\end{align*} 
with the approximation becoming valid for $(at)^{1 / d} \gg 1$. Here, $\Gamma$ denotes the Gamma function. For  $\gamma^2 > 0$ we find
\begin{align*}
&\sum\limits_{m = 0}^\infty \frac{e^{- \lambda'_m t}}{\Vert v_m \Vert^2} \approx \frac{1}{(at)^{2 - 1 / d}}\sum\limits_{m = 0}^\infty \frac{e^{- \frac{1}{y(m)^d}}}{\gamma^2 \frac{y^{2d}}{(at)^2}} \frac{1}{(at)^{1 / d}} \\
&\;\; \approx \frac{1}{(at)^{2 - 1 / d}} \int\limits_{0}^\infty  \frac{e^{- \frac{1}{x^d}}}{\gamma^2 x^{2d}} \text{d}x
= \frac{\Gamma(2 - \frac{1}{d})}{\gamma^2 d} \frac{1}{(at)^{2 - 1 / d}},
\end{align*}
which is again only valid when $(at)^{1/d} \gg 1$. The final relaxation behaviors on long timescales are stated in the main text.

\section{Special solutions} \label{sec_special_solutions}

In the following, we will derive a few more exact solutions for the Pick function. We begin with the Lorentzian distribution as defined in \eref{eq_rates_lorenzian_h}. Note, in the case of $2 bc \in \mathbb{Z}$ the cross-relaxation rates are not distinct from each other and the Pick function is not directly applicable. In this case, the irrelevant eigenvalues given by the degenerate cross-relaxation rates have to be treated beforehand. The analysis is then very similar to that shown in the following, which is valid for $2 bc \notin \mathbb{Z}$.  
The Pick function can be rewritten as
\begin{align*}
    \frac{f(\lambda)}{a} &\equiv \gamma - \frac{1}{z} - \sum\limits_{h = -\infty}^\infty \frac{b^2}{(z - 1)b^2 - (k + b c)^2} \\
    &= \gamma - \frac{1}{z} - \frac{\pi b}{2 \sqrt{z - 1}}\Big[\cot\left(\pi b c + \pi b \sqrt{z - 1}\right) \\ & \qquad \qquad \qquad \qquad \quad \quad  - \cot\left(\pi b c - \pi b \sqrt{z - 1}\right) \Big].
\end{align*}
Moving from the first to the second line requires $z > 1$, which becomes valid for $\lambda \gtrsim \lambda_{0}$. Unfortunately, when taking the derivative, the divergent terms cannot be removed simultaneously, except for the symmetric cases $bc \in \{0, 1 / 4, 1 / 2\}$.  For instance, for $bc = 1 / 4$ we obtain 
\begin{align*}
    \frac{f(\lambda)}{a} = \gamma - \frac{1}{z} + \frac{\pi b}{\sqrt{z - 1}} \tan(2 \pi b \sqrt{z - 1})
\end{align*}
and
\begin{align*}
    \Vert v_m \Vert^2 &= 1 + \frac{1}{2}\frac{ (\gamma z_m - 1) z_m}{(z_m - 1)^2} + \frac{\pi^2 b^2 z_m^2}{z_m - 1} + (\gamma z_m - 1)^2.
\end{align*}
Note, when expressing this solution as a continuous function $z(m) = (m / 2 + 1/4  + \delta(m) / 2)^2 = (m  + 1/2  + \delta(m))^2 / 4$, the TLSs with negative and positive detuning have to be considered.
To compare this result with the one from the distribution defined in \eref{eq_rates_quadratic},
the parameter $a$ must be scaled by a factor of 4.

Next, we discuss the distribution defined in \eref{eq_gamma_d}, which can be solved analytically for integer values $d > 1$. We only present the illustrative calculations of $d = 3$ and $d = 4$. For $d = 4$ it holds
\begin{align*}
\frac{f(\lambda)}{a} &= \gamma - \frac{1}{z} - \sum\limits_{k = 1}^\infty \frac{1}{z - k^4} \\
                     &= \gamma - \frac{1}{z} - \frac{1}{4z^{3/4}}\sum\limits_{k = 1}^\infty \frac{1}{\sqrt[4]{z} + k} + \frac{1}{\sqrt[4]{z} - k}\\[-5pt]
  & \quad\qquad\qquad\qquad\qquad+\frac{1}{\sqrt[4]{z} + ik} + \frac{1}{\sqrt[4]{z} - ik} \\
  &= \gamma - \frac{1}{2z} - \frac{\pi \text{cot}\left(\pi \sqrt[4]{z} \right)}{4z^{3/4}} - \frac{\pi\text{coth}\left(\pi \sqrt[4]{z} \right)}{4z^{3/4}}.
\end{align*}
Expressing the oscillating cotangent through regular terms via the Pick function yields
\begin{align*}
\Vert v_m \Vert^2 &= \frac{3}{8} + \frac{\pi z_m^{1 / 4}}{4}\text{coth}\left(\pi \sqrt[4]{z_m} \right) + \frac{\pi^2 z_m^{2/4}}{8} - \frac{\gamma z_m}{4} \\ & \qquad - \frac{\pi \gamma z_m^{5/4}}{2}\text{coth}\left(\pi \sqrt[4]{z_m} \right) + \gamma^2 z_m^2.
\end{align*}
Representing this solution again as a continuous function $z(m) = (m + 1/2  + \delta(m))^4$, we obtain the expected behavior in the limit $m \rightarrow \infty$ and we find $\beta = \pi^2 / 8$. 
In general, when $d$ is even, $f$ contains $d / 2$ cotangent functions with complex coefficients. Rewriting the oscillating cotangent on the real axis using the Pick function gives a continuous function for $\Vert v_m \Vert^2$.

In principle, we can use the same strategy for $d = 3$:
\begin{align*}
&\frac{f(\lambda)}{a} = \gamma - \frac{1}{z} - \sum\limits_{k = 1}^\infty \frac{1}{z - k^3} \\
&\quad = \dots - \frac{1}{3z^{2/3}}\sum\limits_{k = 1}^\infty \frac{1}{\sqrt[3]{z} - k} + \frac{e^{\frac{2 \pi i}{3}}}{\sqrt[3]{z} e^{\frac{2 \pi i}{3}} - k} + \frac{e^{- \frac{2 \pi i}{3}}}{\sqrt[3]{z}e^{-\frac{2 \pi i}{3}} - k} \\
&\quad = \dots - \frac{1}{3z^{2/3}} \Big[\psi_0 \left(1 -\sqrt[3]{z}\right) + e^{\frac{2 \pi i}{3}}  \psi_0 \left(1 - \sqrt[3]{z} e^{\frac{2 \pi i}{3}}\right) \\
& \hspace{3cm}+ e^{- \frac{2 \pi i}{3}} \psi_0 \left(1 - \sqrt[3]{z} e^{- \frac{2 \pi i}{3}}\right)\Big] \\
&\quad = \dots - \frac{1}{3z^{2/3}} \Big[\pi\cot(\pi\sqrt[3]{z}) + \psi_0(\sqrt[3]{z})  \\
& \hspace{1cm} - e^{\frac{2 \pi i}{6}}  \psi_0 (\sqrt[3]{z} e^{\frac{2 \pi i}{6}}) - e^{- \frac{2 \pi i}{6}} \psi_0 (\sqrt[3]{z} e^{- \frac{2 \pi i}{6}}) - \frac{2}{\sqrt[3]{z}} \Big] \\
&\quad = \gamma - \frac{1}{3 z} - \frac{1}{3z^{2/3}} \Big[\pi\cot(\pi\sqrt[3]{z}) + \frac{\pi}{\sqrt{3}} + \frac{1}{2z^{1/3}} + I(\sqrt[3]{z})\Big] \\
&\quad = \gamma - \frac{1}{2 z} - \frac{\pi \cot(\pi\sqrt[3]{z})}{3z^{2/3}}  + \frac{\pi}{\sqrt{27}z^{2/3}} + \frac{I(\sqrt[3]{z})}{3z^{2/3}},
\end{align*}
\newpage \noindent where $\psi_0$ is the digamma function and $I(\sqrt[3]{z}) = \mathcal{O}(1 / z^{4/3})$ is an integral expression that vanishes continuously and sufficiently fast as $z \rightarrow \infty$.
The integral expression enters as
\begin{align*}
\psi(z) = \ln(z) - \frac{1}{2z} - \int\limits_0^\infty \left(\frac{1}{2} - \frac{1}{t} + \frac{1}{e^{-t} - 1}\right) e^{-zt} \text{d}t,
\end{align*}
valid for $\text{Re}(z) > 0$. By taking the derivative and removing the divergent cotangent, one obtains $\beta = 4 \pi^2 / 27$. The procedure essentially works for any integer $d \geq 2$. If one is only interested in the closed form expression for $\Vert v_m \Vert^2$, one may only use the reflection formula once to introduce the cotangent, via $\psi_0(1- \sqrt[d]{z}) = \pi \cot(\pi \sqrt[d]{z}) + \psi_0(\sqrt[d]{z})$. In this case, the poles on the negative real axis are canceled with $\psi_0(\sqrt[d]{z})$.
\vspace{1.8cm}

%

\end{document}